\documentclass[%
 reprint,
nofootinbib,
 amsmath,amssymb,
 aps,
 prc,
colorlinks=true, linkcolor=blue, citecolor=blue, urlcolor=blue,
floatfix,
]{revtex4-2}

\usepackage[utf8]{inputenc}
\usepackage{babel}
\usepackage{graphicx}

\usepackage{tikz}
\usepackage{etoolbox}
\usepackage{array}          
\usetikzlibrary{positioning,arrows.meta,shapes.multipart,fit,backgrounds,calc}

\ifcsname bdnoalgo\endcsname\else\usepackage{algpseudocode}\fi

\ifcsname bd@loaded\endcsname\expandafter \fi
\makeatletter
\def\bd@loaded{v1.2}

\definecolor{sunIndigo}{HTML}{364B9A}   
\definecolor{sunBlue}  {HTML}{4A7BB7}   
\definecolor{sunSky}   {HTML}{6EA6CD}   
\definecolor{sunSand}  {HTML}{E8B84B}   
\definecolor{sunAmber} {HTML}{FDB366}   
\definecolor{sunOrange}{HTML}{F67E4B}   
\definecolor{sunRed}   {HTML}{DD3D2D}   
\definecolor{sunMaroon}{HTML}{A50026}   
\definecolor{sunGrey}  {HTML}{6B6B6B}   

\colorlet{bdInk}{black!75}              
\colorlet{bdFlow}{sunOrange!78!black}   
\colorlet{bdFrame}{black!55}            

\newlength{\bdunit}      
\newlength{\bdrule}      

\tikzset{
  bd@base/.style = {
    font          = \footnotesize,
    text          = bdInk,
    >             = {Stealth[length=2.4mm,width=2.0mm]},
    line cap      = round,
  },
}

\newcommand{\bd@userstyle}[2]{\tikzset{#1/.style/.expand once=#2}}

\newcommand{\bd@rigidmath}{%
  \thinmuskip=3mu \medmuskip=4mu \thickmuskip=5mu\relax}
\newcommand{\bd@flush}[1]{%
  \ifdefstring{#1}{left}{\def#1{flush left}}{}%
  \ifdefstring{#1}{center}{\def#1{flush center}}{}%
  \ifdefstring{#1}{right}{\def#1{flush right}}{}%
}

\newcommand{\bd@zeroleft}{\leftskip=1\leftskip\relax}
\newcommand{\bd@ragright}{%
  \parindent=0pt
  \rightskip=0pt plus 1fil
  \parfillskip=0pt plus 1fil\relax}
\newcommand{\bdraggedright}{\bd@zeroleft\bd@ragright}

\newcommand{\bdgap}[1][5pt]{\par\vskip#1\relax}

\pgfkeys{/bd/box/.cd,
  color/.store in      = \bd@bcol,
  width/.store in      = \bd@bwd,
  minheight/.store in  = \bd@bmh,
  anchor/.store in     = \bd@banc,
  align/.store in      = \bd@bal,
  titlealign/.store in = \bd@btal,
  font/.store in       = \bd@bfont,
  titlefont/.store in  = \bd@btfont,
  tikz/.store in       = \bd@btikz,
  deliver/.store in    = \bd@bdel,
  deliverlabel/.store in = \bd@bdellab,
  ratio/.store in      = \bd@bratio,   
  gap/.store in        = \bd@bgap,     
  divider/.store in    = \bd@bdiv,     
}
\newcommand{\bd@boxdefaults}{%
  \pgfkeys{/bd/box/.cd,
    color=sunSky, width=4.6cm, minheight=0pt, anchor=center,
    align=left, titlealign=left, font=\footnotesize,
    titlefont=\footnotesize\bfseries, tikz={},
    deliver={}, deliverlabel={Deliver.},
    ratio=0.5, gap=8pt, divider=true}%
}
\bd@boxdefaults

\tikzset{
  bd@box/.style = {
    draw            = \bd@bcol!58!black,
    line width      = \bdrule,
    rounded corners = 2pt,
    text width      = \bd@bwd,
    minimum height  = \bd@bmh,
    anchor          = \bd@banc,
    align           = \bd@bal,
    inner sep       = 4pt,
    font            = \bd@bfont,
  },
  bd@split/.style = {
    bd@box,
    rectangle split, rectangle split parts=2,
    rectangle split part align = {\bd@btal,\bd@bal},
    rectangle split part fill  = {\bd@bcol!35!white,\bd@bcol!11!white},
    rectangle split draw splits = true,
  },
  bd@solo/.style = {
    bd@box,
    fill = \bd@bcol!22!white,
    align = \bd@btal,
  },
  bd@splitdel/.style = {
    bd@box,
    rectangle split, rectangle split parts=3,
    rectangle split part align = {\bd@btal,\bd@bal,\bd@bal},
    rectangle split part fill  = {\bd@bcol!35!white,\bd@bcol!11!white,
                                  \bd@bcol!22!white},
    rectangle split draw splits = true,
  },
  bd@solodel/.style = {
    bd@box,
    rectangle split, rectangle split parts=2,
    rectangle split part align = {\bd@btal,\bd@bal},
    rectangle split part fill  = {\bd@bcol!35!white,\bd@bcol!22!white},
    rectangle split draw splits = true,
  },
}

\newcommand{\bd@delpart}{%
  {\bd@btfont\color{\bd@bcol!62!black}\bd@bdellab}\enspace\bd@bdel}

\newcommand{\bdbox}[5][]{%
  \bd@boxdefaults\pgfkeys{/bd/box/.cd,#1}%
  \bd@flush{\bd@bal}\bd@flush{\bd@btal}%
  \bd@userstyle{bd@ubox}{\bd@btikz}%
  \expandafter\ifblank\expandafter{\bd@bdel}%
    {\def\bd@hasdel{0}}{\def\bd@hasdel{1}}%
  \ifblank{#5}{%
    \ifnum\bd@hasdel=0\relax
      \node[bd@solo,bd@ubox] (#2) at (#3)
        {{\bd@btfont\color{\bd@bcol!62!black}#4}};%
    \else
      \node[bd@solodel,bd@ubox] (#2) at (#3)
        {\nodepart{one}{\bd@btfont\color{\bd@bcol!62!black}#4}%
         \nodepart{two}{\bd@delpart}};%
    \fi
  }{%
    \ifnum\bd@hasdel=0\relax
      \node[bd@split,bd@ubox] (#2) at (#3)
        {\nodepart{one}{\bd@btfont\color{\bd@bcol!62!black}#4}%
         \nodepart{two}{#5}};%
    \else
      \node[bd@splitdel,bd@ubox] (#2) at (#3)
        {\nodepart{one}{\bd@btfont\color{\bd@bcol!62!black}#4}%
         \nodepart{two}{#5}%
         \nodepart{three}{\bd@delpart}};%
    \fi
  }%
}

\pgfkeys{/bd/arrow/.cd,
  color/.store in     = \bd@acol,
  width/.store in     = \bd@awd,
  route/.store in     = \bd@aroute,
  bend/.store in      = \bd@abend,
  label/.store in     = \bd@alab,
  pos/.store in       = \bd@apos,
  labelside/.store in = \bd@aside,
  style/.store in     = \bd@asty,
  offset/.store in    = \bd@aoff,
  tikz/.store in      = \bd@atikz,
}
\newcommand{\bd@arrdefaults}{%
  \pgfkeys{/bd/arrow/.cd,
    color=bdFlow, width=0.9pt, route=straight, bend=25, label={},
    pos=0.5, labelside=on, style={}, offset=-8mm, tikz={}}%
}
\bd@arrdefaults

\tikzset{
  bd@arr/.style = {
    -{Stealth[length=2.4mm,width=2.0mm]},
    draw            = \bd@acol,
    line width      = \bd@awd,
    rounded corners = 1.5pt,
  },
  bd@arrlab/.style = {
    font      = \scriptsize,
    text      = bdInk,
    inner sep = 1.6pt,
  },
}

\newcommand{\bd@arrgo}[3]{%
  \draw[bd@arr,bd@uarrsty,bd@uarr] (#1) #2
    node[bd@arrlab,bd@ularr,pos=\bd@apos]{\bd@alab} (#3);%
}

\newcommand{\bdarrow}[3][]{%
  \bd@arrdefaults\pgfkeys{/bd/arrow/.cd,#1}%
  \expandafter\ifblank\expandafter{\bd@alab}%
    {\def\bd@labopts{inner sep=0pt}}%
    {\ifdefstring{\bd@aside}{on}%
       {\def\bd@labopts{fill=white}}%
       {\edef\bd@labopts{\bd@aside}}}%
  \bd@userstyle{bd@ularr}{\bd@labopts}%
  \bd@userstyle{bd@uarr}{\bd@atikz}%
  \bd@userstyle{bd@uarrsty}{\bd@asty}%
  \ifdefstring{\bd@aroute}{hv}{\bd@arrgo{#2}{-|}{#3}}{%
   \ifdefstring{\bd@aroute}{vh}{\bd@arrgo{#2}{|-}{#3}}{%
    \ifdefstring{\bd@aroute}{hvh}%
      {\bd@arrgo{#2}{-- ++(\bd@aoff,0) |-}{#3}}{%
     \ifdefstring{\bd@aroute}{vhv}%
       {\bd@arrgo{#2}{-- ++(0,\bd@aoff) -|}{#3}}{%
      \ifdefstring{\bd@aroute}{bend}%
        {\bd@arrgo{#2}{to[bend left=\bd@abend]}{#3}}%
        {\bd@arrgo{#2}{--}{#3}}}}}}%
}

\pgfkeys{/bd/group/.cd,
  color/.store in     = \bd@gcol,
  fill/.store in      = \bd@gfill,
  style/.store in     = \bd@gsty,
  pad/.store in       = \bd@gpad,
  labelpos/.store in  = \bd@glpos,
  labelfont/.store in = \bd@glfont,
  tikz/.store in      = \bd@gtikz,
}
\newcommand{\bd@grpdefaults}{%
  \pgfkeys{/bd/group/.cd,
    color=bdFrame, fill=none, style=dashed, pad=7pt,
    labelpos=nw, labelfont=\scriptsize\itshape, tikz={}}%
}
\bd@grpdefaults

\tikzset{
  bd@grp/.style = {
    draw            = \bd@gcol,
    fill            = \bd@gfill,
    line width      = 0.5pt,
    dash pattern    = on 2.2pt off 1.8pt,
    rounded corners = 3pt,
    inner sep       = \bd@gpad,
  },
  bd@grplab/.style = {
    font      = \bd@glfont,
    text      = \bd@gcol,
    fill      = white,
    inner sep = 2pt,
  },
}

\newcommand{\bdgroup}[4][]{%
  \bd@grpdefaults\pgfkeys{/bd/group/.cd,#1}%
  \bd@userstyle{bd@ugrp}{\bd@gtikz}%
  \bd@userstyle{bd@ugrpsty}{\bd@gsty}%
  \begin{pgfonlayer}{background}
    \node[bd@grp,bd@ugrpsty,bd@ugrp, fit={#3}] (#2) {};
  \end{pgfonlayer}
  \expandafter\ifblank\expandafter{\bd@glpos}{}{%
    \ifdefstring{\bd@glpos}{nw}
      {\node[bd@grplab, anchor=west]  at ([xshift=8pt]#2.north west) {#4};}{}%
    \ifdefstring{\bd@glpos}{n}
      {\node[bd@grplab]               at (#2.north)                  {#4};}{}%
    \ifdefstring{\bd@glpos}{sw}
      {\node[bd@grplab, anchor=west]  at ([xshift=8pt]#2.south west) {#4};}{}%
    \ifdefstring{\bd@glpos}{s}
      {\node[bd@grplab]               at (#2.south)                  {#4};}{}%
    \ifdefstring{\bd@glpos}{w}
      {\node[bd@grplab, rotate=90]    at (#2.west)                   {#4};}{}%
  }%
}

\pgfkeys{/bd/note/.cd,
  width/.store in  = \bd@nwd,
  anchor/.store in = \bd@nanc,
  align/.store in  = \bd@nal,
  font/.store in   = \bd@nfont,
  color/.store in  = \bd@ncol,
  tikz/.store in   = \bd@ntikz,
}
\newcommand{\bd@notedefaults}{%
  \pgfkeys{/bd/note/.cd,
    width=8cm, anchor=north west, align=left, font=\scriptsize,
    color=bdInk, tikz={}}%
}
\bd@notedefaults

\newcommand{\bdnote}[4][]{%
  \bd@notedefaults\pgfkeys{/bd/note/.cd,#1}%
  \bd@flush{\bd@nal}%
  \bd@userstyle{bd@unote}{\bd@ntikz}%
  \node[text width=\bd@nwd, anchor=\bd@nanc, align=\bd@nal,
        font=\bd@nfont, text=\bd@ncol, inner sep=0pt, bd@unote]
       (#2) at (#3) {#4};%
}

\pgfkeys{/bd/env/.cd,
  fit/.store in  = \bd@efit,
  unit/.store in = \bd@eunit,
  grid/.store in = \bd@egrid,
  tikz/.store in = \bd@etikz,
}
\newcommand{\bd@envdefaults}{%
  \pgfkeys{/bd/env/.cd, fit={}, unit=\bdunit, grid={}, tikz={}}%
}
\bd@envdefaults

\newsavebox{\bd@sbox}
\newlength{\bd@hlen}
\newcommand{\bd@sboxheight}{%
  \setlength{\bd@hlen}{\ht\bd@sbox}%
  \addtolength{\bd@hlen}{\dp\bd@sbox}%
  \edef\bd@hpt{\the\bd@hlen}}

\def\bd@parsegrid#1,#2,#3,#4\@nil{%
  \def\bd@gxa{#1}\def\bd@gya{#2}\def\bd@gxb{#3}\def\bd@gyb{#4}}

\newcommand{\bd@drawgrid}{%
  \expandafter\bd@parsegrid\bd@egrid\@nil
  \begin{pgfonlayer}{background}
    \draw[step=1, draw=black!14, line width=0.2pt]
      (\bd@gxa,\bd@gya) grid (\bd@gxb,\bd@gyb);
    \draw[draw=black!30, line width=0.4pt]
      (\bd@gxa,0) -- (\bd@gxb,0)  (0,\bd@gya) -- (0,\bd@gyb);
    \foreach \bd@i in {\bd@gxa,...,\bd@gxb}{%
      \node[font=\tiny, text=black!45, inner sep=1pt, anchor=south]
        at (\bd@i,\bd@gyb) {\bd@i};}
    \foreach \bd@j in {\bd@gya,...,\bd@gyb}{%
      \node[font=\tiny, text=black!45, inner sep=1pt, anchor=east]
        at (\bd@gxa,\bd@j) {\bd@j};}
  \end{pgfonlayer}
}

\newenvironment{blockdiagram}[1][]{%
  \bd@envdefaults\pgfkeys{/bd/env/.cd,#1}%
  \bd@userstyle{bd@uenv}{\bd@etikz}%
  \begin{lrbox}{\bd@sbox}%
  \bd@rigidmath
  \begin{tikzpicture}[x=\bd@eunit, y=\bd@eunit, bd@base, bd@uenv]%
  \expandafter\ifblank\expandafter{\bd@egrid}{}{\bd@drawgrid}%
}{%
  \end{tikzpicture}%
  \end{lrbox}%
  \edef\bd@wpt{\the\wd\bd@sbox}%
  \pgfmathparse{\bd@wpt/28.4527}\edef\bd@wcm{\pgfmathresult}%
  \bd@sboxheight
  \pgfmathparse{\bd@hpt/28.4527}\edef\bd@hcm{\pgfmathresult}%
  \typeout{[blockdiag] natural size:
    \bd@wcm cm x \bd@hcm cm (\bd@wpt\space x \bd@hpt)}%
  \expandafter\ifblank\expandafter{\bd@efit}%
    {\usebox{\bd@sbox}}%
    {\resizebox{\bd@efit}{!}{\usebox{\bd@sbox}}}%
}

\newcounter{bdregion}

\pgfkeys{/bd/axis/.cd,
  color/.store in    = \bd@xcol,
  end/.store in      = \bd@xend,
  var/.store in      = \bd@xvar,
  ticklen/.store in  = \bd@xtick,
  numerals/.store in = \bd@xnum,
  font/.store in     = \bd@xfont,
  tikz/.store in     = \bd@xtikz,
}
\newcommand{\bd@axisdefaults}{%
  \pgfkeys{/bd/axis/.cd,
    color=sunSky, end={}, var={$\omega$}, ticklen=0.14,
    numerals=roman, font=\scriptsize, tikz={}}%
}
\bd@axisdefaults

\newcommand{\bdregionaxis}[2][]{%
  \bd@axisdefaults\pgfkeys{/bd/axis/.cd,#1}%
  \xdef\bd@xmax{0}%
  \foreach \bd@p/\bd@l in {#2}{%
    \pgfmathsetmacro{\bd@xt}{max(\bd@xmax,\bd@p)}\xdef\bd@xmax{\bd@xt}}%
  \expandafter\ifblank\expandafter{\bd@xend}%
    {\pgfmathsetmacro{\bd@xe}{\bd@xmax+0.6}}%
    {\edef\bd@xe{\bd@xend}}%
  \ifdefstring{\bd@xnum}{arabic}%
    {\def\bd@regionnum{\arabic{bdregion}}}%
    {\def\bd@regionnum{\Roman{bdregion}}}%
  \bd@userstyle{bd@uaxis}{\bd@xtikz}%
  \begin{tikzpicture}[x=1cm, y=1cm, font=\bd@xfont,
      shape=rectangle, fill=none, inner sep=1pt,
      every node/.style={shape=rectangle, fill=none, draw=none,
                         inner sep=1.5pt, text=bdInk},
      bd@uaxis]
    \draw[line width=0.6pt, draw=\bd@xcol!60!black, fill=\bd@xcol!60!black,
          -{Stealth[length=1.7mm,width=1.4mm]}] (0,0) -- (\bd@xe,0);
    \expandafter\ifblank\expandafter{\bd@xvar}{}%
      {\node[anchor=west] at (\bd@xe,0) {\bd@xvar};}%
    \foreach \bd@p/\bd@l in {#2}{%
      \draw[line width=0.5pt, draw=\bd@xcol!60!black]
        (\bd@p,0) -- (\bd@p,-\bd@xtick);
      \node[anchor=north] at (\bd@p,-\bd@xtick) {\bd@l};}%
    \ifdefstring{\bd@xnum}{none}{}{%
      \foreach \bd@p/\bd@l [count=\bd@i, remember=\bd@p as \bd@q] in {#2}{%
        \ifnum\bd@i>1\relax
          \setcounter{bdregion}{\bd@i}\addtocounter{bdregion}{-1}%
          \node[anchor=south, font=\bd@xfont\bfseries,
                text=\bd@xcol!62!black]
            at ({0.5*(\bd@p+\bd@q)},0) {\bd@regionnum};
        \fi}}%
  \end{tikzpicture}%
}

\newcounter{bd@enumdepth}

\newcommand{\bdenumcounter}{bdenum\romannumeral\value{bd@enumdepth}}

\newenvironment{bdenumerate}[1][\arabic{\bdenumcounter}.]{%
  \ifnum\value{bd@enumdepth}>3\relax
    \PackageError{blockdiag}{bdenumerate nested too deeply (max 4)}{}%
  \fi
  \stepcounter{bd@enumdepth}%
  \expandafter\def\csname bd@enumlab\romannumeral\value{bd@enumdepth}\endcsname{#1}%
  \bd@zeroleft                      
  \begin{list}{{\footnotesize
      \csname bd@enumlab\romannumeral\value{bd@enumdepth}\endcsname}}{%
    \usecounter{\bdenumcounter}%
    \setlength{\topsep}{0pt}\setlength{\partopsep}{0pt}%
    \setlength{\parsep}{0pt}\setlength{\itemsep}{3pt}%
    \setlength{\leftmargin}{1.6em}\setlength{\labelwidth}{1.3em}%
    \setlength{\labelsep}{0.3em}\setlength{\rightmargin}{0pt}%
    \setlength{\listparindent}{0pt}}%
  \bd@ragright                      
}{\end{list}\addtocounter{bd@enumdepth}{-1}}

\ifcsname bdnoalgo\endcsname\else

\newcounter{bdalgo}

\pgfkeys{/bd/algo/.cd,
  width/.store in  = \bd@awd,
  color/.store in  = \bd@acol,
  font/.store in   = \bd@afont,
  name/.store in   = \bd@aname,
  number/.store in = \bd@anum,
  label/.store in  = \bd@alabel,
  lineno/.store in = \bd@alineno,
}
\newcommand{\bd@algodefaults}{%
  \pgfkeys{/bd/algo/.cd,
    width=\linewidth, color=sunSky, font=\scriptsize,
    name=Algorithm, number=true, label={}, lineno=1}%
}
\bd@algodefaults

\newenvironment{bdalgo}[2][]{%
  \bd@algodefaults\pgfkeys{/bd/algo/.cd,#1}%
  \begin{minipage}[t]{\bd@awd}%
    \bdraggedright\bd@afont
    \setlength{\parskip}{0pt}%
    {\color{\bd@acol!58!black}\hrule height 0.7pt}%
    \vskip 2.5pt
    \ifdefstring{\bd@anum}{false}{%
      {\bfseries\color{\bd@acol!62!black}#2}\par
    }{%
      \refstepcounter{bdalgo}%
      \expandafter\ifblank\expandafter{\bd@alabel}{}%
        {\expandafter\label\expandafter{\bd@alabel}}%
      {\bfseries\color{\bd@acol!62!black}\bd@aname~\thebdalgo}%
      \ifblank{#2}{}{\;\,{\color{\bd@acol!62!black}#2}}\par
    }%
    \vskip 2.5pt
    {\color{\bd@acol!58!black}\hrule height 0.4pt}%
    \vskip 2pt
    \ifdefstring{\bd@alineno}{0}%
      {\begin{algorithmic}}%
      {\begin{algorithmic}[\bd@alineno]}%
}{%
    \end{algorithmic}%
    \vskip 1pt
    {\color{\bd@acol!58!black}\hrule height 0.7pt}%
  \end{minipage}%
}

\algrenewcommand{\alglinenumber}[1]{{\tiny\color{bdInk}#1:}}

\fi   

\newlength{\bd@savail}\newlength{\bd@slw}\newlength{\bd@srw}

\newcommand{\bdsplitbox}[6][]{%
  \bd@boxdefaults\pgfkeys{/bd/box/.cd,#1}%
  \setlength{\bd@savail}{\bd@bwd}%
  \addtolength{\bd@savail}{-2\bd@bgap}%
  \addtolength{\bd@savail}{-1pt}%
  \setlength{\bd@slw}{\bd@bratio\bd@savail}%
  \setlength{\bd@srw}{\bd@savail}\addtolength{\bd@srw}{-\bd@slw}%
  \ifdefstring{\bd@bdiv}{false}{%
    \bdbox[#1]{#2}{#3}{#4}{%
      \setlength{\extrarowheight}{0pt}\renewcommand{\arraystretch}{1}%
      \begin{tabular}{@{}c@{\hskip 2\bd@bgap}c@{}}
        \begin{minipage}[t]{\bd@slw}\bdraggedright #5\end{minipage} &
        \begin{minipage}[t]{\bd@srw}\bdraggedright #6\end{minipage}
      \end{tabular}}%
  }{%
    \bdbox[#1]{#2}{#3}{#4}{%
      \setlength{\extrarowheight}{0pt}\renewcommand{\arraystretch}{1}%
      \begin{tabular}{@{}c@{\hskip\bd@bgap}%
          !{\color{\bd@bcol!58!black}\vrule width 0.4pt}%
          @{\hskip\bd@bgap}c@{}}
        \begin{minipage}[t]{\bd@slw}\bdraggedright #5\end{minipage} &
        \begin{minipage}[t]{\bd@srw}\bdraggedright #6\end{minipage}
      \end{tabular}}%
  }%
}

\newcommand{\bdpalette}{%
  \begin{blockdiagram}
    \foreach \bd@c/\bd@n [count=\bd@k from 0] in {%
      sunIndigo/sunIndigo, sunBlue/sunBlue, sunSky/sunSky,
      sunSand/sunSand, sunAmber/sunAmber, sunOrange/sunOrange,
      sunRed/sunRed, sunMaroon/sunMaroon, sunGrey/sunGrey}{%
      \bdbox[color=\bd@c, width=2.4cm, titlealign=center, align=center]
            {sw\bd@k}{\bd@k*2.9,0}{\bd@n}{fill \& accent};}
  \end{blockdiagram}%
}

\pgfkeys{/bd/conn/.cd,
  color/.store in = \bd@ccol,
  size/.store in  = \bd@csize,
  font/.store in  = \bd@cfont,
  tikz/.store in  = \bd@ctikz,
}
\newcommand{\bd@conndefaults}{%
  \pgfkeys{/bd/conn/.cd,
    color=bdInk, size=5.5mm, font=\footnotesize\bfseries, tikz={}}%
}
\bd@conndefaults

\newcommand{\bdconnector}[4][]{%
  \bd@conndefaults\pgfkeys{/bd/conn/.cd,#1}%
  \bd@userstyle{bd@uconn}{\bd@ctikz}%
  \node[circle, draw=\bd@ccol, line width=0.6pt, fill=white,
        minimum size=\bd@csize, inner sep=0pt,
        font=\bd@cfont, text=\bd@ccol, bd@uconn]
       (#2) at (#3) {#4};%
}

\makeatother
\usepackage{dcolumn}
\usepackage{bm}
\makeatletter
\def\linenumberpar@ltx{\ifvmode\@@@par\else\ifinner\@@@par\else\@linenumberpar\fi\fi}
\makeatother
\usepackage{newtxtext}
\usepackage{newtxmath}
\usepackage{adjustbox}
\usepackage{subcaption}
\usepackage{amsfonts,amsmath, amssymb}
\usepackage{physics}
\usepackage{siunitx}
\usepackage{lipsum}
\usepackage{xcolor}
\usepackage{multirow}
\usepackage{pgfplots}
\usepackage{isotope}
\pgfplotsset{compat=1.18}
\usepackage{xspace}
\usepackage[T1]{fontenc} 
\usepackage{xparse}
\usepackage[normalem]{ulem}

\usepackage{caption}
\usepackage{newfloat}

\DeclareFloatingEnvironment[name=Algorithm, fileext=loa, placement=tbp]{algorithm}
\algrenewcommand{\algorithmicrequire}{\textbf{Input:}}
\algrenewcommand{\algorithmicensure}{\textbf{Output:}}
\renewcommand{\thesubfigure}{(\alph{subfigure})}
\newsavebox{\figoverlaybox}
\tikzset{
  figcite/.style={anchor=base west, inner sep=0,
    font=\sffamily\fontsize{\figcitefs pt}{\figcitefs pt}\selectfont},
  figlabel/.style={anchor=north east,
    inner xsep=\figlabelxsep pt, inner ysep=\figlabelysep pt,
    font=\fontsize{\figlabelfs pt}{\figlabelfs pt}\selectfont},
}
\newcommand{\figoverlay}[3]{%
  \sbox{\figoverlaybox}{\includegraphics[#1]{#2}}%
  \pgfmathsetmacro{\figcitefs}{0.037*\wd\figoverlaybox}%
  \pgfmathsetmacro{\figlabelfs}{0.045*\wd\figoverlaybox}%
  \pgfmathsetmacro{\figlabelxsep}{0.9*\figlabelfs}
  \pgfmathsetmacro{\figlabelysep}{0.5*\figlabelfs}%
  \begin{tikzpicture}
    \node[anchor=south west, inner sep=0] (img) at (0,0) {\usebox{\figoverlaybox}};
    \begin{scope}[x={(img.south east)}, y={(img.north west)}]
      #3
    \end{scope}
  \end{tikzpicture}%
}
\newcommand{\figcitenode}[3]{\node[figcite] at (#1,#2) {\cite{#3}};}
\newcommand{\figlabelnode}[2]{\node[figlabel] at (#1,#2) {\thesubfigure};}

\NewDocumentCommand{\labeledgraphics}{O{width=\linewidth} m m m}{%
  \phantomsubcaption
  \figoverlay{#1}{#2}{\figlabelnode{#3}{#4}}%
  \ignorespaces
}

\ExplSyntaxOn
\tl_new:N \l__figov_cites_tl
\NewDocumentCommand{\citedgraphics}{O{width=\linewidth} m m m m}
  {
    \tl_set:Nn \l__figov_cites_tl { \figcitenode {#3} {#4} {#5} }
    \__figov_cited_next:nn {#1} {#2}
  }
\cs_new_protected:Npn \__figov_cited_next:nn #1#2
  {
    \peek_meaning_ignore_spaces:NTF \c_group_begin_token
      { \__figov_cited_add:nnnnn {#1} {#2} }
      { \figoverlay {#1} {#2} { \l__figov_cites_tl } }
  }
\cs_new_protected:Npn \__figov_cited_add:nnnnn #1#2#3#4#5
  {
    \tl_put_right:Nn \l__figov_cites_tl { \figcitenode {#3} {#4} {#5} }
    \__figov_cited_next:nn {#1} {#2}
  }

\NewDocumentCommand{\citedlabeledgraphics}{O{width=\linewidth} m m m m}
  {
    \tl_set:Nn \l__figov_cites_tl { \figcitenode {#3} {#4} {#5} }
    \__figov_citedlabeled_next:nnnn {#1} {#2}
  }
\cs_new_protected:Npn \__figov_citedlabeled_next:nnnn #1#2#3#4
  {
    \peek_meaning_ignore_spaces:NTF \c_group_begin_token
      { \__figov_citedlabeled_add:nnnnn {#1} {#2} {#3} {#4} }
      {
        \phantomsubcaption
        \figoverlay {#1} {#2} { \l__figov_cites_tl \figlabelnode {#3} {#4} }
      }
  }
\cs_new_protected:Npn \__figov_citedlabeled_add:nnnnn #1#2#3#4#5
  {
    \tl_put_right:Nn \l__figov_cites_tl { \figcitenode {#3} {#4} {#5} }
    \__figov_citedlabeled_next:nnnn {#1} {#2}
  }
\ExplSyntaxOff

\newcommand{\nuwro}{\textsc{NuWro}\xspace}
\newcommand{\ewro}{\texorpdfstring{\textsc{$e$Wro}}{eWro}\xspace}
\newcommand{\ewronet}{\texorpdfstring{\textsc{$e$WroNet}}{eWroNet}\xspace}
\newcommand{\neut}{\textsc{NEUT}\xspace}
\newcommand{\genie}{\textsc{GENIE}\xspace}
\newcommand{\egenie}{\ensuremath{e}-\textsc{GENIE}\xspace}
\newcommand{\gibuu}{\textsc{GiBUU}\xspace}
\newcommand{\achilles}{\textsc{ACHILLES}\xspace}

\newcommand{\ttok}{T2K\xspace}
\newcommand{\hyperk}{Hyper-Kamiokande\xspace}

\newcommand{\dune}{DUNE\xspace}

\newcommand{\clas}{CLAS\xspace}
\newcommand{\efnu}{$e4\nu$\xspace}

\newcommand{\eezeropi}{$(e, e')_{0\pi}$\xspace}

\newcommand{\qe}{QE\xspace}
\newcommand{\mec}{MEC\xspace}
\newcommand{\res}{RES\xspace}
\newcommand{\spp}{SPP\xspace}
\newcommand{\dis}{DIS\xspace}

\newcommand{\fsi}{FSI\xspace}
\newcommand{\inel}{INEL\xspace}
\newcommand{\INEL}{\mathrm{INEL}\xspace}

\newcommand{\xF}[1]{\sigma''^{\mathrm{F1F2}}_{#1}\xspace}
\newcommand{\xE}[1]{\sigma''^{\ewro}_{#1}\xspace}
\newcommand{\xD}[1]{\sigma''^{\mathrm{DNN}}_{#1}\xspace}
\newcommand{\xT}[1]{\hat{\sigma}''_{#1}\xspace}
\newcommand{\QE}{\mathrm{QE}\xspace}
\newcommand{\MEC}{\mathrm{MEC}\xspace}
\newcommand{\SPP}{\mathrm{SPP}\xspace}
\newcommand{\inc}{\mathrm{inc}\xspace}
\newcommand{\maxx}{\mathrm{max}\xspace}
\newcommand{\minn}{\mathrm{min}\xspace}
\newcommand{\qeerec}{E^{\mathrm{rec}}_{\QE}\xspace}

\usepackage{orcidlink}

\usepackage{hyperref}

\definecolor{JanRed}{rgb}{0.8, 0.0, 0.0}

\definecolor{KrzysztofYellow}{rgb}{1.000, 0.761, 0.039}

\definecolor{HemantBlue}{rgb}{0.0, 0.2, 0.7}

\begin{document}

\preprint{APS/123-QED}

\title{Incorporating deep neural networks into lepton-nucleus interaction modeling}


\author{Hemant Prasad \orcidlink{0009-0003-1897-9616}}
\email{hemant.prasad@phd.uwr.edu.pl}

\author{Jan T. Sobczyk \orcidlink{0000-0003-4991-2919}}
\email{jan.sobczyk@uwr.edu.pl}


\author{Krzysztof M. Graczyk \orcidlink{0000-0002-0038-6340}}

\author{Beata E. Kowal \orcidlink{0000-0003-3646-1653}}

\author{Rwik Dharmapal Banerjee \orcidlink{0000-0003-3639-7532}}

\author{J. Luis Bonilla \orcidlink{0009-0009-3240-1494} }

\affiliation{   Institute of Theoretical Physics,
                University of Wroc{\l}aw,
                Plac Maxa Borna 9, 50-204,
                Wroc{\l}aw, Poland
            }

\date{\today}

\begin{abstract}
We present a scheme to incorporate precise inclusive cross-section predictions from a deep neural network trained on electron–carbon scattering data [Kowal~\textit{et al.} \href{https://doi.org/10.1103/ydvc-2567}{Phys. Rev. C \textbf{112}, 055504 (2025)}] into a Monte Carlo event generator. The inclusive prediction is decomposed into independent exclusive interaction channels using the generator’s theoretical models, and the resulting events are reweighted accordingly. We apply this scheme to investigate nuclear effects in inclusive and semi-exclusive electron-scattering measurements.
\end{abstract}

\maketitle

\section{\label{sec:Introduction}INTRODUCTION}


\par The next generation of long-baseline neutrino oscillation experiments, \dune~\cite{JINST.15.T08008, arXiv.2002.03005} and \hyperk~\cite{Prog.Theor.Exp.Phys.2015.053C02}, aims to measure precisely the values of the oscillation parameters, and in particular to determine the
charge-parity-violating phase in the lepton sector. Because the oscillation probability varies
rapidly with neutrino energy while accelerator-based beams are broad, the incident energy must be reconstructed from the detected final-state particles. Any reconstruction relies on a reliable description of the nuclear target, the elementary interaction mechanisms, and the
propagation of the produced hadrons through the nuclear medium. As a result, modeling neutrino-nucleus interactions is among the leading sources of systematic uncertainty in the oscillation analyses~\cite{Prog.Part.Nucl.Phys.100.1}.

\par A possible way to understand uncertainties in neutrino-nucleus interactions is to explore electron scattering off the nucleus since electron and neutrino interactions share many similarities. Electrons interact with nucleons through vector currents, while neutrinos interact via both vector and axial-vector currents. Both interactions are modeled in the impulse approximation and assume the same nuclear initial state and final-state interactions (\fsi). In electron scattering experiments, the beam is monoenergetic, and the energy-momentum transfer is known exactly on an event-by-event basis. Any model of neutrino interactions should also be able to reproduce electron scattering data.  For carbon, precise inclusive cross-section measurements have accumulated over five decades and span a broad kinematic region~\cite{Phys.Rev.C.53.2248, Phys.Rev.Lett.82.2056, Bagdasaryan:1988hp, Phys.Rev.Lett.61.400, Nucl.Phys.A.402.515, Phys.Rev.C.98.014617, Phys.Rev.C.48.1849, Phys.Rev.Lett.105.212502, Phys.Rev.C.35.1063, Sealock:1989nx, Phys.Rev.C.9.2230, Few.Body.Syst.65.78, Phys.Rev.D.49.4348}. Accordingly, reproducing these data in electron-scattering mode is a minimal requirement for neutrino Monte Carlo (MC) event generators such as \nuwro~\cite{NuWro_GitHub_2025}, \neut~\cite{Hayato}, \genie~\cite{Nucl.Instrum.Meth.A.614.87}, \gibuu~\cite{Phys.Rept.512.1}, and \achilles~\cite{PhysRevD.107.033007}, particularly when aiming to reduce systematic uncertainties to the sub-percent level.


\par Several neutrino MC event generators have therefore been extended to electron scattering. The \egenie covers quasielastic (\qe) scattering, resonance production (\res), meson-exchange currents (\mec), and deep-inelastic scattering (\dis), see Refs.~\cite{PhysRevD.103.113003, Khachatryan2021}. In \nuwro, electron scattering is simulated by the \ewro module, which describes \qe scattering and single-pion production (\spp)~\cite{_muda_2015}; its \qe description has recently been improved with spectral function developments~\cite{PhysRevD.109.073004, ankowski2025spectralfunctionapproachnuwro}. In \neut, electron scattering was first implemented for \qe scattering~\cite{wreo32248} and has recently been extended to \spp based on the dynamical coupled-channels model~\cite{PhysRevD.111.033006}. \gibuu provides a unified theory and transport framework capable of simulating electron-nucleus scattering~\cite{Phys.Rept.512.1, Phys.Rev.D.109.033008}, whereas \achilles currently includes only \qe scattering~\cite{PhysRevD.107.033007} in its electron-scattering mode. Considerable effort has been devoted to validating the implementations of theoretical models in the electron-scattering mode of MC event generators~\cite{_muda_2015, PhysRevD.111.033006, wreo32248, PhysRevD.103.113003, Eur.Phys.J.C.79.293, Phys.Rev.D.91.033005, Phys.Rev.C.100.045503, Khachatryan2021}. Although existing theoretical models describe some kinematic regions remarkably well, they do not consistently reproduce the experimental data across all kinematics, indicating the need for further improvements indeed. The study of the \clas and \efnu collaborations at Jefferson Lab (JLab)~\cite{Khachatryan2021}, based on comparisons with \egenie~\cite{PhysRevD.103.113003}, concluded that current models need to be improved to meet the requirements of next-generation, high-precision experiments.

\par Alongside theory-driven models, machine learning techniques have increasingly been used to derive model-independent representations of nuclear cross sections directly from experimental data~\cite{bslk-gxr7}. In recent years, approaches based on deep neural networks (DNNs) have gained prominence in lepton-nucleus scattering, enabling data-driven models to predict inclusive electron-nucleus cross sections~\cite{Phys.Rev.C.107.065501, Phys.Rev.C.110.025501,Phys.Rev.C.112.055504,Phys.Rev.Lett.052502,Phys.Lett.B.859.139142}. Al Hammal~\textit{et al.}~\cite{Phys.Rev.C.107.065501} trained a single neural network on data for nuclei from $^{4}$He to $^{59}$Ni and tested its predictions on nuclei excluded from the training, such as $^{40}$Ca, $^{40}$Ar, and $^{48}$Ti. In contrast, the DNN model introduced by Kowal~\textit{et al.}~\cite{Phys.Rev.C.110.025501,Phys.Rev.C.112.055504} was trained using a bootstrap data set constructed from existing electron-carbon scattering measurements, producing an empirical fit to inclusive cross-section data. The choice of carbon was motivated by the extensive kinematic coverage available in the data set~\cite{Phys.Rev.C.53.2248, Phys.Rev.Lett.82.2056, Bagdasaryan:1988hp, Phys.Rev.Lett.61.400, Nucl.Phys.A.402.515, Phys.Rev.C.98.014617, Phys.Rev.C.48.1849, Phys.Rev.Lett.105.212502, Phys.Rev.C.35.1063, Sealock:1989nx, Phys.Rev.C.9.2230, Few.Body.Syst.65.78, Phys.Rev.D.49.4348}. Furthermore, Graczyk~\textit{et al.}~\cite{Phys.Rev.Lett.052502} demonstrated that DNNs trained on carbon data could be adapted, through transfer learning techniques~\cite{zhuang2020comprehensivesurveytransferlearning, J.Big.Data.3.9}, to predict cross sections for a variety of other nuclear targets, provided the relevant kinematic coverage exists for the chosen target. An alternative strategy, proposed by Sobczyk~\textit{et al.}~\cite{Phys.Lett.B.859.139142}, involves training artificial neural networks on longitudinal and transverse response functions to predict inclusive cross sections across different nuclear targets.

\par The DNN models~\cite{Phys.Rev.C.107.065501, Phys.Rev.C.110.025501,Phys.Rev.C.112.055504,Phys.Rev.Lett.052502,Phys.Lett.B.859.139142} have shown strong effectiveness in predicting inclusive cross sections. However, their predictions are currently limited to the inclusive cross section and do not convey how the cross section is distributed among exclusive  interaction channels, such as \qe scattering, \mec, etc., nor do they provide information on the kinematics of the final-state particles. Both aspects are essential for MC event generators, which require each event to be fully specified and thus necessitate a nuclear model. The approach described in this article integrates both methodologies: it decomposes the inclusive DNN prediction into exclusive interaction channels and employs theoretical models to generate the corresponding final states.

\par In this article, we present a framework to decompose the inclusive DNN model  by Kowal~\textit{et al.}~\cite{Phys.Rev.C.110.025501, Phys.Rev.C.112.055504} into distinct exclusive interaction channels and incorporate it into the \ewro MC event generator~\cite{_muda_2015} by reweighting the existing theoretical models for \qe and \spp to generate the kinematics of the final-state particles. For \mec, we construct an approximate \mec model using the parametrization of Ref.~\cite{bosted} to infer the inclusive component, while the exclusive component is described using the hadronic model~\cite{PhysRevD.111.036032}. We call the resulting framework \ewronet, which provides precise inclusive predictions from the DNN and the exclusive predictions from the theoretical models.

The article is organized as follows: Section~\ref{sec:Methodology} outlines the scheme and the data used to benchmark it. Sections~\ref{sec:nuwro}--\ref{subsec:implementation_ewro} describe \nuwro with its electron-scattering module \ewro, the DNN model, and \ewronet, respectively. Section~\ref{sec:Results} compares \ewronet with electron-scattering data, and Sect.~\ref{sec:discussion} discusses the approach's limitations and possible extensions. Appendices~\ref{sec:MEC} and~\ref{sec:delta-peak-width-derivation} present the approximate \mec model and derive formulas used in Section~\ref{subsec:implementation_ewro} for the position and width of the $\Delta(1232)$ peak. 

\section{\label{sec:Methodology}METHODOLOGY}

This study aims to develop a MC event generator that accurately reproduces the inclusive cross section while providing predictions for hadronic observables. This approach fixes the inclusive cross section using a data-driven model, the DNN~\cite{Phys.Rev.C.110.025501, Phys.Rev.C.112.055504}, enabling more reliable estimates of the associated systematic uncertainties. We achieve this by developing the \ewronet scheme, which is based on two independent ingredients:
\begin{enumerate}
  \item The \ewro MC event generator, incorporating the theoretical models of \qe~\cite{ankowski2025spectralfunctionapproachnuwro} and \spp~\cite{_muda_2015}, which predict the kinematics of the exclusive final-state particles. For \mec, we use the approximate model described in Appendix~\ref{sec:MEC}.
  \item The DNN parametrization of the inclusive cross section for a carbon target~\cite{Phys.Rev.C.110.025501, Phys.Rev.C.112.055504}.
\end{enumerate}
We present the scheme for carbon target, but it can be extended to other nuclear targets in a similar manner. The first ingredient provides the nuclear models needed to generate complete MC events for electron scattering on carbon. It also determines each exclusive channel's fractional contribution to the inclusive cross section. The second ingredient provides the value of the inclusive cross section. Using each channel's fractional contribution, we decompose the DNN inclusive cross section into exclusive channels. We then determine the required readjustment to each event weight such that the sample reproduces the DNN prediction for the inclusive cross section across the kinematic range covered by the scheme. We describe each component: the \ewro generator, the DNN model, and their integration within the \ewro MC generator, in detail in Sections~\ref{sec:nuwro}--\ref{subsec:implementation_ewro}.

We then perform a consistency check of the scheme by comparing with the inclusive cross sections from Sealock~\textit{et al.}~\cite{Sealock:1989nx} at three kinematic settings. We chose these settings because they lie in the energy range relevant for the \ttok and \hyperk experiments and are closest to the kinematics of the \clas and \efnu measurements at JLab (Sect.~\ref{subsection:jlabdata}). We then benchmark our proposed scheme against inclusive and semi-exclusive measurements from the \clas and \efnu collaborations~\cite{Khachatryan2021}. Both data sets are described below.

\subsection{\label{subsection:inclusivedata} Inclusive data}

The DNN model~\cite{Phys.Rev.C.110.025501, Phys.Rev.C.112.055504} used in this study was trained on the inclusive electron-carbon cross sections of Refs.~\cite{Phys.Rev.C.53.2248, Phys.Rev.Lett.82.2056, Bagdasaryan:1988hp, Phys.Rev.Lett.61.400, Nucl.Phys.A.402.515, Phys.Rev.C.98.014617, Phys.Rev.C.48.1849, Phys.Rev.Lett.105.212502, Phys.Rev.C.35.1063, Sealock:1989nx, Phys.Rev.C.9.2230, Few.Body.Syst.65.78, Phys.Rev.D.49.4348}, which span $0\le Q^2\le10$~GeV$^{2}$. The data are given as double-differential cross sections, $\sigma''\equiv d^{2}\sigma/d\omega\,d\Omega$, as functions of the energy transfer $\omega$ at various beam energies $E$ and scattering angles $\theta$. Although the DNN describes the full data set well (see Figs.~6--10 of Ref.~\cite{Phys.Rev.C.112.055504}), we demonstrate the consistency of the proposed scheme on a subset of it: the measurements performed at the Stanford Linear Accelerator Center (SLAC) by Sealock~\textit{et al.}~\cite{Sealock:1989nx} at $\theta=37.5^\circ$ and $E=0.961$, $1.108$, and $1.299$~GeV. 

\subsection{\label{subsection:jlabdata}\clas and \efnu measurements}

The \clas and \efnu collaborations measured electron scattering on $^{4}$He, $^{12}$C, and $^{56}$Fe at JLab with beam energies $E=1.159$, $2.257$, and $4.453$~GeV, and tested interaction models by reconstructing the known energy of the monoenergetic beam with the methods used in neutrino-oscillation analyses~\cite{Khachatryan2021}. Scattered electrons were detected at $15^\circ\le\theta\le45^\circ$ with energies $E'$ above $0.4$, $0.55$, and $1.1$~GeV, respectively. Hadrons with momenta above $0.15$--$0.3$~GeV$/c$ within $10^{\circ}\lesssim\theta_h\le140^{\circ}$, and photons with energies $E_\gamma\ge0.3$~GeV, were detected as well. Events with the scattered electron and the detected $\pi^{-}$ were further restricted by fiducial cuts, see Eqs.~(12)--(14) of Ref.~\cite{Khachatryan2021}, to maximize signal efficiency.

From these data, a pionless sample, $(e,e')_{0\pi}$, was constructed, containing events with any number of nucleons and no charged pions or photons above the detection thresholds. The sample was analyzed for the reconstructed energy defined as:
\begin{equation}
    \label{eqn:Erec}
    E^{\rm rec}_{\QE} = \frac{2M\epsilon + 2ME' - m_l^2}{2\left(M - E' + |\mathbf{k}'|\cos\theta\right)},
\end{equation}
where $M$ is the nucleon mass; $m_l$, $E'$, $|\mathbf{k}'|$, and $\theta$ are the mass, energy, momentum, and scattering angle of the outgoing lepton; and $\epsilon=21$~MeV is the average nucleon separation energy. Additionally, a sample with exactly one detected proton, $(e,e'p)_{1p0\pi}$, was obtained by subtracting the contributions of events with additional undetected protons above the threshold. Since the electron-nucleon cross section is much more forward peaked than the neutrino-nucleon one, all events were weighted by $Q^4$. In addition, the collaborations extracted the inclusive cross section at $E=1.159$~GeV and $\theta=37.5^\circ$ from the electrons detected in a single CLAS sector within $36^\circ\le\theta\le39^\circ$, normalized to the numbers of incident electrons and target nuclei; see Extended Data Fig.~3 of Ref.~\cite{Khachatryan2021}.

\section{\label{sec:nuwro}\nuwro  MC Event Generator}

\nuwro~\cite{NuWro_GitHub_2025} is an MC event generator for lepton-nucleus scattering that has been under continuous development at the University of Wroc{\l}aw since 2004, with accelerator-based neutrino experiments as its principal application. It is applicable to lepton energies from a few hundred MeV up to several hundred GeV. At the primary vertex, the generator implements \qe scattering, quasielastic hyperon production, \mec, \res, and \dis, each available for charged-current and neutral-current processes. Several descriptions of the initial state of the target nucleus are available: the hole spectral function (SF)~\cite{Nucl.Phys.A.505.267, Nucl.Phys.A.579.493}, an effective potential with density and momentum dependence~\cite{Juszczak:2005wk}, and the Fermi gas in its global or local form.

\subsection{\label{subsection:ewro} \ewro}

\nuwro includes a submodule called \ewro~\cite{_muda_2015}, whose primary purpose is to validate interaction models against electron-scattering data. Its core functionality largely mirrors that of \nuwro; it uses a factorization scheme and can generate weighted or unweighted events for a specified target, either from an incident flux or from a monoenergetic beam of energy $E$. Unlike \nuwro, \ewro generates events in a specified final lepton angular region defined by the cosine of the scattering angle, $\cos\theta$, and the cone width, $\delta\cos\theta$. Currently, \ewro supports electron scattering via \qe~\cite{ PhysRevD.109.073004, ankowski2025spectralfunctionapproachnuwro} and \spp in the $\Delta$ resonance region~\cite{Phys.Rev.D.76.033005, _muda_2015}, but it contains no \mec model. Since the \mec contribution is needed to describe the dip region between the \qe and $\Delta$ peaks, we construct an approximate \mec model, described in Appendix~\ref{sec:MEC}. In all channels, the hadrons produced at the primary vertex propagate through the nucleus via the same intranuclear cascade~\cite{Golan:2012wx}, in which particles move along straight-line trajectories between collisions and the nucleon-nucleon cross sections include in-medium modifications~\cite{PhysRevC.45.791}.

\subsubsection{\label{subsubsec:QESF} SF description of QE}

\qe scattering is treated within the impulse approximation, in which the electron scatters off a single bound nucleon and the remaining $(A-1)$ nucleons act as spectators~\cite{Rev.Mod.Phys.80.189}. The electron-nucleon cross section is obtained in the Llewellyn Smith formalism~\cite{Phys.Rept.3.261}, retaining only the vector current, with the nucleon form factors of the BBBA05 parametrization~\cite{Nucl.Phys.B.Proc.Suppl.159.127}. The target is described by the hole SF $P(\mathbf{p},E)$~\cite{Nucl.Phys.A.505.267, Nucl.Phys.A.579.493}, which gives the probability of removing a nucleon of momentum $\mathbf{p}$ from the nuclear ground state, leaving the residual system with excitation energy $E$. For carbon, we employ the SF of Ref.~\cite{Phys.Rev.C.110.054612}, which combines the shell structure measured in $(e,e'p)$ experiments at Saclay and NIKHEF with the contribution of short-range correlations obtained within the local-density approximation~\cite{Nucl.Phys.A.579.493}. Pauli blocking is implemented by rejecting events in which the momentum of the knocked-out nucleon lies below the average Fermi momentum. The Coulomb distortion of the electron and the recoil of the residual nucleus are also taken into account~\cite{Phys.Rev.D.91.033005}. The \fsi of the struck nucleon with the spectator system are included through the convolution scheme, which broadens the \qe peak and shifts its position through the real part of the optical potential. With these ingredients, the SF approach reproduces the inclusive electron-carbon data in the \qe region with remarkable precision. A detailed description of recent improvements to the \qe channel, along with comparisons to inclusive electron-scattering measurements, appears in Ref.~\cite{ ankowski2025spectralfunctionapproachnuwro}.

\subsubsection{\label{subsubsec:SPP} SPP}

In \ewro, \spp on a bound nucleon,
\[ e+N\rightarrow e'+N'+\pi,\]
is described in the four isospin channels
\begin{eqnarray*}
  e+p\rightarrow e'+p+\pi^0,\\
  e+p\rightarrow e'+n+\pi^+,\\
  e+n\rightarrow e'+n+\pi^0,\\
  e+n\rightarrow e'+p+\pi^-.
\end{eqnarray*}
The model includes the $\Delta(1232)$ resonance together with the nonresonant background terms required by chiral symmetry, derived from the SU(2) nonlinear $\sigma$ model of Ref.~\cite{Phys.Rev.D.76.033005}; the pion-pole term, being purely axial, does not contribute to electroproduction. Several parametrizations of the electromagnetic $N\rightarrow\Delta(1232)$ transition form factors are available~\cite{Phys.Rev.D.90.093001, AIP.Conf.Proc.1680.020013, Phys.Rev.D.71.074003}. We use the one from Ref.~\cite{Phys.Rev.D.71.074003}. 
The model was validated against inclusive electron-scattering measurements in Ref.~\cite{_muda_2015}. The target nucleon is drawn from the local Fermi gas, and the outgoing pion and nucleon kinematics are generated exclusively. Since heavier resonances are not included, the model is applicable approximately only up to the right half-maximum of the $\Delta(1232)$ peak.

\section{ \label{subsec:DNN} The DNN model\texorpdfstring{~\cite{Phys.Rev.C.110.025501, Phys.Rev.C.112.055504}}{}}
\label{sec:dnn}

In this study, we use the DNN model of the inclusive electron-carbon cross section developed by Kowal~\textit{et al.}~\cite{Phys.Rev.C.110.025501, Phys.Rev.C.112.055504}. The original model~\cite{Phys.Rev.C.110.025501} was
trained on eleven independent data sets~\cite{Phys.Rev.C.53.2248, Phys.Rev.Lett.82.2056, Bagdasaryan:1988hp, Phys.Rev.Lett.61.400,
Nucl.Phys.A.402.515, Phys.Rev.C.98.014617, Phys.Rev.C.48.1849, Phys.Rev.Lett.105.212502, Phys.Rev.C.35.1063, Sealock:1989nx,
Phys.Rev.C.9.2230}, which cover the \qe peak, the resonance region, and the onset of \dis. Later on, the model was re-optimized~\cite{Phys.Rev.C.112.055504} by incorporating the new measurements~\cite{Few.Body.Syst.65.78, Phys.Rev.D.49.4348} performed at $Q^2=0.8$~GeV$^2$ at a scattering angle of $70^\circ$, covering both the elastic and inelastic regions. In this work, we employ the re-optimized model from Ref.~\cite{Phys.Rev.C.112.055504}.

At lower energy transfers, the inclusive cross section receives contributions from elastic scattering off the whole
nucleus and from the excitation of the giant dipole resonance and of discrete nuclear states. As the data in this region are too
scarce to constrain these processes, they were removed from the training~\cite{Phys.Rev.C.110.025501, Phys.Rev.C.112.055504} by applying the kinematic cut $\omega>\omega_{\min}$, where $\omega_\minn$ is given as:
\begin{equation}
   \omega_{\min} = \frac{M_A E_x + 0.5 E_x^2 + E^2(1-\cos\theta)}{M_A + E(1-\cos\theta)},
   \label{eq:omegamin}
\end{equation}
where $M_A$ is the mass of the target nucleus and $E_x=26$~MeV is the excitation energy. We note that $\omega_{\min}\approx E_x+t_A$, where $t_A$ is the nucleus's recoil energy. The DNN predictions are therefore meaningful
only when $\omega>\omega_{\min}$ is satisfied. Below $\omega_\minn$, we rely on a scaled response from the \qe SF model of \ewro, see Sect.~\ref{subsec:implementation_ewro}, to match smoothly with the DNN output.

The model is an ensemble of $N_{\rm nn}=50$ networks, each trained on a distinct bootstrap replica of the data. Each network takes as input the beam energy $E$ (in GeV), the energy transfer $\omega$ (in GeV), the scattering angle $\theta$, as well as
$\cos\theta$ and the squared four-momentum transfer $Q^{2}=4E(E-\omega)\sin^{2}(\theta/2)$ (in GeV$^2$). Only the first three inputs are independent; the remaining two can be determined from them. Adding them only accelerated the optimization procedure~\cite{Phys.Rev.C.110.025501}. We provide them to the network because they are part of the architecture inputs.
The output layer is a sigmoid function, $y\in(0,1)$, which is related to the double-differential cross section
$\sigma''^{\rm DNN}$ through
\begin{equation}
  \sigma''^{\rm DNN} = \frac{10^{9}\,\sigma_{\rm Mott}}{E\cos(\theta/2)}\,y,
  \qquad
  \text{where}\;\sigma_{\rm Mott} = \frac{\alpha^{2}\cos^{2}(\theta/2)}{4E^{2}\sin^{4}(\theta/2)},
  \label{eq:dnn-output}
\end{equation}
with $\sigma''$ expressed in units of nb/(GeV\,sr).

The cross-section prediction of the model $\xD{\inc}(\omega)$ is the ensemble mean,  and is related to the  prediction of the $r$th network at fixed $(E,\theta)$, as:
\begin{equation}
  \label{eqn:dnn-curve}
  \xD{\inc}(\omega) = \frac{1}{N_{\rm nn}}\sum_{r=1}^{N_{\rm nn}}\xD{\inc,r}(\omega).
\end{equation}
The standard deviation of the ensemble expresses the predictive uncertainty of the model,
\begin{equation}
  \label{eqn:uncertainty-dnn-curve}
  \Delta\xD{\inc}(\omega) = \left[\frac{1}{N_{\rm nn}}\sum_{r=1}^{N_{\rm nn}}
  \left(\xD{\inc,r}(\omega) - \xD{\inc}(\omega)\right)^{2}\right]^{1/2}.
\end{equation}
A detailed account of the architecture, the training procedure, and the validation of the model is given in Refs.~\cite{Phys.Rev.C.110.025501, Phys.Rev.C.112.055504}. The performance of DNN predictions is depicted in Figs.~6--10 of Ref.~\cite{Phys.Rev.C.112.055504}, which show that the DNN reproduces the inclusive cross section precisely, indicating that DNN predictions can serve as a proxy for inclusive electron-carbon scattering data at any chosen $E$ and $\theta$ where no measurements are available. This sets up our argument of incorporating the DNN model~\cite{Phys.Rev.C.110.025501, Phys.Rev.C.112.055504}, provided that we decompose it into exclusive channels, for precise comparisons.

\section{\label{subsec:implementation_ewro}\ewronet}

\begin{figure*}[!t]
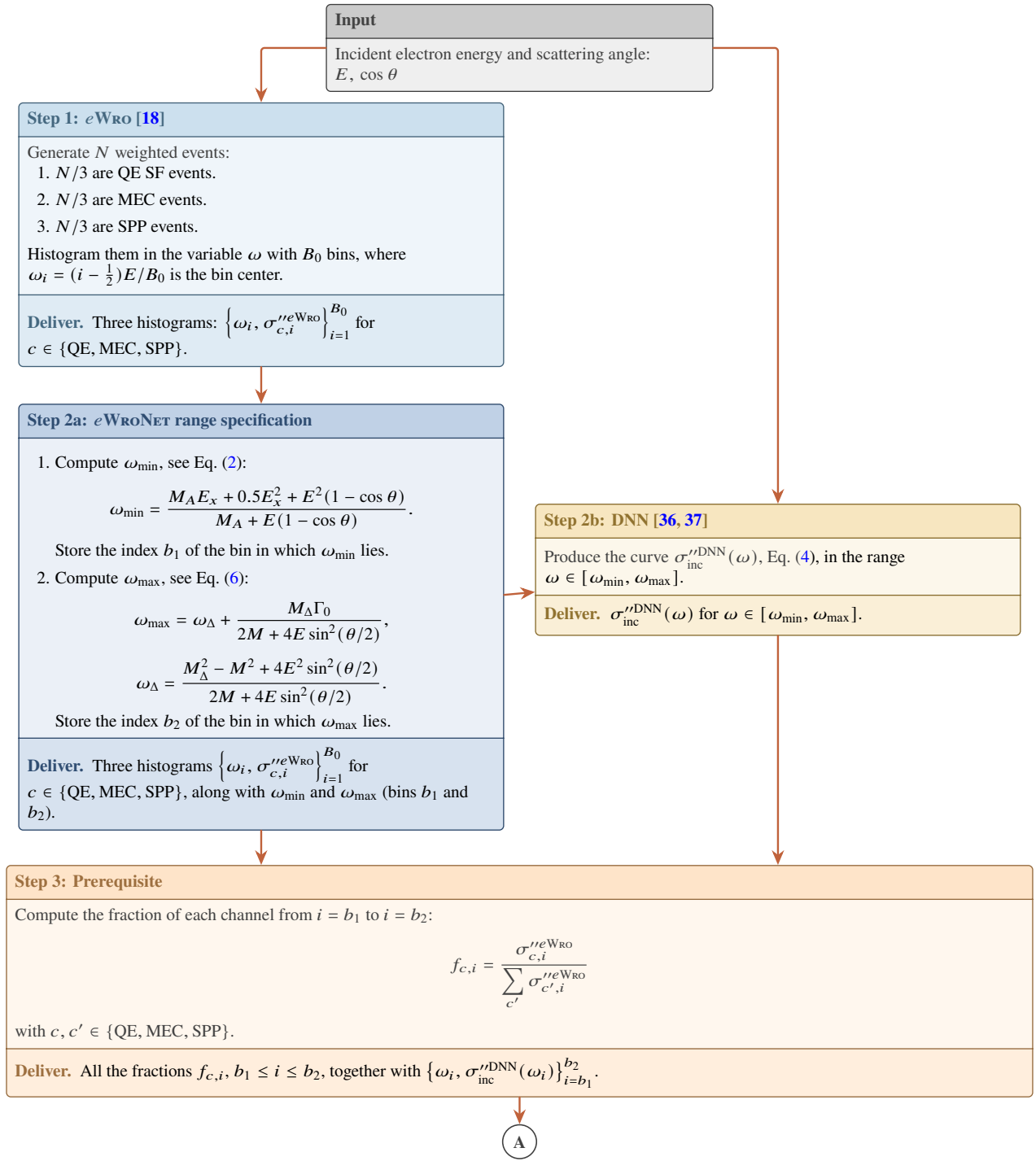

    \centering
    \begin{blockdiagram}

        \bdbox[color=sunGrey, width=6cm, anchor=north]{input}{0,0}{Input}
              {Incident electron energy and scattering angle: $E,\,\cos\theta$}

        \bdbox[color=sunSky, width=7.6cm, anchor=north,
               deliver={Three histograms: $\left\{\omega_i,\xE{c,i}\right\}_{i=1}^{B_0}$
                        for $c\in\{\QE,\MEC, \SPP\}$.}]%
              {ew}{-4.2,-1.6}{Step~1: \ewro~\cite{NuWro_GitHub_2025}}%
              {Generate $N$ weighted events:
               \begin{bdenumerate}
                 \item $N/3$ are \qe SF events.
                 \item $N/3$ are \mec events.
                 \item $N/3$ are \spp events.
               \end{bdenumerate}\bdgap[3pt]
               Histogram them in the variable $\omega$ with $B_0$ bins, where
               $\omega_i=(i-\frac{1}{2})E/B_0$ is the bin center.}


        \bdbox[color=sunBlue, width=7.6cm, anchor=north,
               deliver={Three histograms $\left\{\omega_i,\xE{c,i}\right\}_{i=1}^{B_0}$
                        for $c\in\{\QE,\MEC,\SPP\}$, along with $\omega_\minn$ and
                        $\omega_\maxx$ (bins $b_1$ and $b_2$).}]%
              {range}{-4.2,-6.5}{Step~2a: \ewronet range specification}%
              {\begin{bdenumerate}
                 \item Compute $\omega_\minn$, see Eq.~(\ref{eq:omegamin}):
                       \[\omega_\minn = \frac{M_AE_x+0.5E_x^2+E^2(1-\cos\theta)}
                                              {M_A+E(1-\cos\theta)}.\]
                       Store the index $b_1$ of the bin in which $\omega_\minn$ lies.
                 \item Compute $\omega_\maxx$, see Eq.~(\ref{eq:omegamax}):
                       \[\omega_\maxx = \omega_\Delta
                         + \frac{M_\Delta\Gamma_0}{2M+4E\sin^2(\theta/2)},\]
                       \[\omega_\Delta = \frac{M_\Delta^2-M^2+4E^2\sin^2(\theta/2)}
                                              {2M+4E\sin^2(\theta/2)}.\]
                       Store the index $b_2$ of the bin in which $\omega_\maxx$ lies.
               \end{bdenumerate}}

        \bdbox[color=sunSand, width=7.6cm,
               deliver={$\xD{\inc}(\omega)$ for $\omega\in[\omega_\minn,\omega_\maxx]$.}]%
              {dnn}{4.2,-9.2}{Step~2b: DNN~\cite{Phys.Rev.C.110.025501, Phys.Rev.C.112.055504}}%
              {Produce the curve $\xD{\inc}(\omega)$, Eq.~(\ref{eqn:dnn-curve}),
               in the range $\omega\in[\omega_\minn,\omega_\maxx]$.}

        \bdbox[color=sunAmber, width=16.4cm, anchor=north,
               deliver={All the fractions $f_{c,i}$, $b_1\le i\le b_2$,
                        together with $\left\{\omega_i, \xD{\inc}(\omega_i)\right\}_{i=b_1}^{b_2}$.}]%
              {prereq}{0,-14.0}{Step~3: Prerequisite}%
              {Compute the fraction of each channel from
                 $i=b_1$ to $i=b_2$:
                 \[f_{c,i} = { \displaystyle \xE{c,i} \over \displaystyle \sum_{c'}\xE{c',i}}\]
                 with $c,c'\in\{\QE,\MEC,\SPP\}$.
                }

        \bdarrow[route=hv]{input.west}{ew.north}
        \bdarrow[route=hv]{input.east}{dnn.north}
        \bdarrow{ew}{range}
        \bdarrow{range}{dnn}
        \bdarrow{range.south}{range.south |- prereq.north}
        \bdarrow{dnn.south}{dnn.south |- prereq.north}

        \bdconnector{connA}{0,-18.5}{A}
        \bdarrow{prereq}{connA}

    \end{blockdiagram}
    \caption{\label{fig:block-diagram}(a) Block diagram of the \ewronet
      algorithm, part one: inputs, \ewro (Step~1), the range
      specification (Step~2a), the DNN (Step~2b), and the prerequisites (Step~3). The flow
      continues at connector~A in part (b) on the next page.}
\end{figure*}

\begin{figure*}[!t]
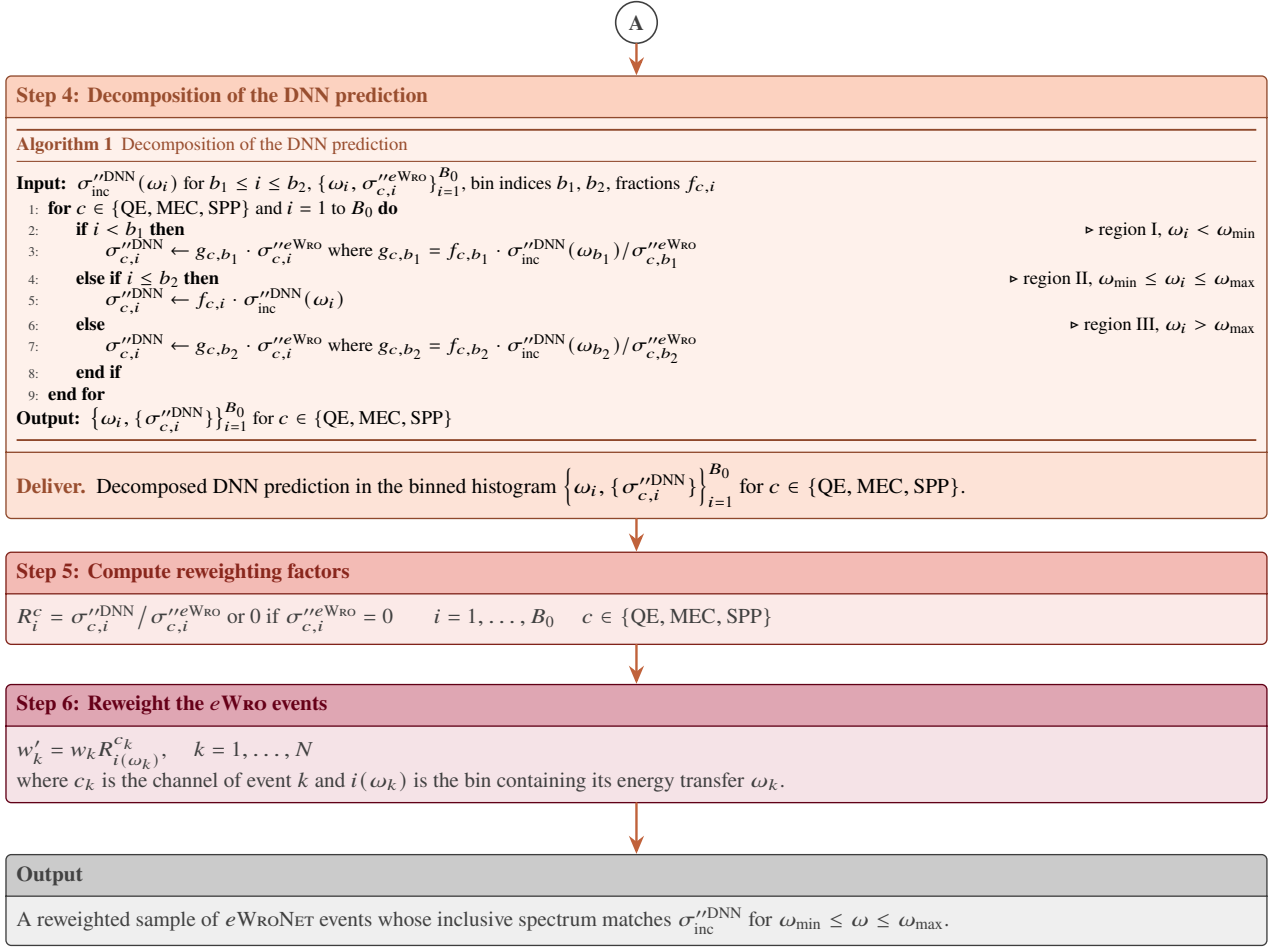

    \ContinuedFloat
    \centering
    \begin{blockdiagram}

        \bdconnector{connA}{0,0}{A}

        \bdbox[color=sunOrange, width=16.4cm, anchor=north,
               deliver={Decomposed DNN prediction in the binned histogram
                        $\left\{\omega_i,\{\xD{c,i}\}\right\}_{i=1}^{B_0}$
                        for $c\in\{\QE,\MEC,\SPP\}$.}]%
              {dec}{0,-0.7}{Step~4: Decomposition of the DNN prediction}%
              {\begin{bdalgo}[color=sunOrange, label=alg:decompose]{Decomposition of the DNN prediction}
                 \Require $\xD{\inc}(\omega_i)$ for $b_1\le i\le b_2$,
                          $\{\omega_i,\xE{c,i}\}_{i=1}^{B_0}$, bin indices $b_1$, $b_2$,
                          fractions $f_{c,i}$
                 \For{$c\in\{\QE,\MEC,\SPP\}$ and $i = 1$ to $B_0$}
                   \If{$i<b_1$} \Comment{region I, $\omega_i<\omega_\minn$}
                     \State $\xD{c,i} \gets g_{c, b_1}\cdot\xE{c, i}$ where $g_{c,b_1} = f_{c,b_1}\cdot\xD{\inc}(\omega_{b_1})/\xE{c,b_1}$
                   \ElsIf{$i\le b_2$} \Comment{region II, $\omega_\minn\le\omega_i\le\omega_\maxx$}
                     \State $\xD{c,i} \gets f_{c,i}\cdot\xD{\inc}(\omega_i)$
                   \Else \Comment{region III, $\omega_i>\omega_\maxx$}
                     \State $\xD{c,i} \gets g_{c, b_2}\cdot\xE{c, i}$ where $g_{c,b_2} = f_{c,b_2}\cdot\xD{\inc}(\omega_{b_2})/\xE{c,b_2}$
                   \EndIf
                 \EndFor
                 \Ensure $\bigl\{\omega_i,\{\xD{c,i}\}\bigr\}_{i=1}^{B_0}$
                         for $c\in\{\QE,\MEC,\SPP\}$
               \end{bdalgo}}

        \bdbox[color=sunRed, width=16.4cm, anchor=north]{s4}{0,-7.0}%
              {Step~5: Compute reweighting factors}%
              {$R^{c}_{i}=\xD{c, i}\big/\xE{c,i}$ or 0 if $\xE{c, i} = 0$ \,\,\,\,\quad
               $i=1,\dots,B_{0}$ \quad $c\in\{\QE,\MEC,\SPP\}$}

        \bdbox[color=sunMaroon, width=16.4cm, anchor=north]{s5}{0,-8.75}%
              {Step~6: Reweight the \ewro events}%
              {$w'_{k}=w_{k}R^{c_{k}}_{i(\omega_{k})}$, \quad $k=1,\dots,N$\\
               where $c_k$ is the channel of event $k$ and $i(\omega_k)$ is the bin containing
               its energy transfer $\omega_k$.}

        \bdbox[color=sunGrey, width=16.4cm, anchor=north]{output}{0,-11}%
              {Output}%
              {A reweighted sample of \ewronet events whose inclusive
               spectrum matches $\xD{\inc}$ for $\omega_\minn\le\omega\le\omega_\maxx$.}

        \bdarrow{connA}{dec}
        \bdarrow{dec}{s4}
        \bdarrow{s4}{s5}
        \bdarrow{s5}{output}

    \end{blockdiagram}
    \caption{\label{fig:block-diagram-b}(Continued.) (b) Part two: Steps
      4--6, and the output of the \ewronet scheme, starting from connector~A of part (a).}
\end{figure*}

The DNN model of Sect.~\ref{sec:dnn} was trained with Python packages. To incorporate it into \ewro, we export the model to the Open Neural Network Exchange (ONNX) format and load it in \ewro with the ONNX Runtime library~\cite{onnxruntime}. For a given beam energy $E$ and scattering angle $\theta$, \ewro is then run together with the DNN. The inclusive DNN prediction is decomposed into exclusive channels to reweight the existing theoretical models, producing complete MC events. We refer to the combined framework as \ewronet. The scheme is general in that it assumes only that models of the exclusive channels are implemented in MC, which then serve as input to the decomposition. For \mec, we developed an approximate model that serves as such an input (Appendix~\ref{sec:MEC}); in the future, it can be replaced by any microscopic calculation that offers better physical insight. Analogously, in the future, one can add heavier resonances and \dis to \ewronet. Below, we describe the steps of the algorithm presented in Fig.~\ref{fig:block-diagram}.

In Step~1, \ewro generates a sufficiently large number of weighted events at the chosen $(E,\cos\theta)$, divided equally among the \qe, \mec, and \spp channels. The \qe events are described within the SF approach (Sect.~\ref{subsubsec:QESF}), the \spp events using the model~\cite{_muda_2015} (Sect.~\ref{subsubsec:SPP}), and the \mec events within the approximate model of Appendix~\ref{sec:MEC}. We histogram the events of each channel $c\in\{\QE,\MEC,\SPP\}$ in $\omega$ with $B_0$ bins, yielding the double-differential cross sections $\xE{c,i}$ at the bin centers $\omega_i$, $i=1,\dots,B_0$, in the same units as the DNN output.

In Step~2a, we determine the range in which the inclusive cross section of \ewronet is given by the DNN prediction. The lower boundary is the cut $\omega_\minn$ of Eq.~(\ref{eq:omegamin}), below which the DNN was not trained. The upper boundary is set by the physics coverage of the \spp model~\cite{_muda_2015}, which includes only the $\Delta(1232)$ resonance and the nonresonant background. At higher energy transfers, higher resonances, multipion production, and \dis contribute, and distributing their strength over the three channels would wrongly assign it to events with $\Delta$-like final states. We therefore place the upper boundary at the right half-maximum of the $\Delta(1232)$ peak,
\begin{equation}
   \omega_\maxx = \omega_{\Delta} + \frac{M_{\Delta}\Gamma_0}{2M + 4E\sin^{2}(\theta/2)},
   \label{eq:omegamax}
\end{equation}
where $\omega_\Delta$ is the position of the peak, see Eq.~(\ref{eqn:peakposition-Delta}), $M$ is the nucleon mass, and $\Gamma_0=0.12$~GeV is the free width of the $\Delta(1232)$. The bin indices containing $\omega_\minn$ and $\omega_\maxx$ are denoted by $b_1$ and $b_2$, respectively. They are crucial for Steps~3 and 4. In Step~2b, the inclusive DNN model produces predictions $\xD{\inc}(\omega)$ within the range $[\omega_\minn,\omega_\maxx]$.

In the following steps we decompose the inclusive DNN prediction into exclusive channels. In Step~3, the fractional contribution of each channel $c\in\{\QE,\MEC,\SPP\}$ to the inclusive cross section of \ewro,
\begin{equation}
  \label{eqn:decomposition-fractions}
  f_{c,i}=\frac{\xE{c,i}}{\sum_{c'}\xE{c',i}}, \qquad i=b_1,\dots,b_2,
\end{equation}
is computed in every bin from $b_1$ to $b_2$. In Step~4 (Algorithm~\ref{alg:decompose}), the DNN prediction is shared among the channels according to these fractions,
\begin{equation}
  \label{eqn:dnn-exclusive-channel-crosssection}
  \xD{c,i} = f_{c,i}\,\xD{\inc}(\omega_i),
\end{equation}
So, by construction, the DNN provides the inclusive cross section, while the theoretical models provide its composition. Below bin $i=b_1$ and above $i=b_2$, each channel follows a scaled version of \ewro's prediction. The expression for the cross section is given as
\begin{equation}
  \xD{c,i} = g_{c,b}\,\xE{c,i}, \qquad g_{c,b}=\frac{f_{c,b}\,\xD{\inc}(\omega_{b})}{\xE{c,b}},
\end{equation}
where $b=b_1$ for $i<b_1$ and $b=b_2$ for $i>b_2$. This ensures a smooth continuation of the inclusive cross section at the boundaries $b_1$ and $b_2$.

In Step~5, the reweighting factors $R^c_i=\xD{c,i}/\xE{c,i}$ are computed, with $R^c_i=0$ whenever $\xE{c,i}=0$. In Step~6, each event $k$ of the sample is assigned the new weight
\begin{equation}
w'_k = w_k\,R^{c_k}_{i(\omega_k)},
\end{equation}
where $c_k$ denotes the interaction channel of the $k$th event and $i(\omega_k)$ is the index of the bin containing its energy transfer $\omega_k$. Because only the weights are modified, the final-state particle kinematics, including their propagation through the intranuclear cascade, remain those of the underlying models. The output is a weighted sample of complete events whose inclusive cross section reproduces the DNN prediction for $\omega_\minn\le\omega\le\omega_\maxx$, and from which unweighted events can be produced in the standard way.

\ewronet produces MC events for a fixed beam energy $E$ and the cosine of the scattering angle $\cosine\theta$. To generate events over an angular range, the range can be divided into $L$ slices of width $\delta\cos\theta$, narrow enough for the cross section to vary little within each slice. \ewronet is then run at the central angle $\theta_l$ of each slice, $l=1,\dots,L$, with the desired number of events, and the resulting samples, each normalized to the cross section of its slice, are merged to form the final sample for the angular range.


\section{\label{sec:Results} RESULTS}

In this section, we compare \ewronet predictions with the electron-scattering data on carbon described in Sect.~\ref{sec:Methodology}. We first test it against the inclusive measurements of Sealock~\textit{et al.}~\cite{Sealock:1989nx} at $\theta=37.5^\circ$ and $E=0.961$, $1.108$, and $1.299$~GeV. As these data were used to train the DNN~\cite{Phys.Rev.C.110.025501, Phys.Rev.C.112.055504}, this test checks the consistency of the scheme rather than its predictive power. We then compare \ewronet with the inclusive and semi-exclusive measurements of the \clas and \efnu collaborations~\cite{Khachatryan2021}. The inclusive cross section was measured at $E=1.159$~GeV and $\theta=37.5^\circ$, close to the setting $E=1.108$~GeV from Ref.~\cite{Sealock:1989nx}, but was not used in the training of the DNN. For semi-exclusive measurements, we restrict ourselves to the pionless sample \eezeropi, for which the reconstructed energy $\qeerec$ is determined solely by the kinematics of the scattered electron, because the approximate \mec model (Appendix~\ref{sec:MEC}) contains no microscopic calculation and therefore does not provide reliable predictions for the outgoing nucleons. 

For the semi-exclusive comparison, we run \ewronet at $E=1.159$~GeV in the angular range $15^\circ\le \theta \le 45^\circ$ by dividing it into $L=100$ equal-width slices in $\cos\theta$, as described in Sect.~\ref{subsec:implementation_ewro}. We then apply the event-selection criteria, see Sect.~\ref{subsection:jlabdata}, rejecting events with charged or neutral pions or photons above the detection thresholds. We then compute the reconstructed energy $\qeerec$ from Eq.~(\ref{eqn:Erec}) for the selected events, and weight each \ewronet event by $Q^4$, as in the experimental analysis.

\subsection{\label{subsec:ewronet-vs-sealock} Comparison with inclusive measurements}

\begin{figure*}[htbp!]
    \centering
    \begin{subfigure}[b]{0.33\linewidth}
      \centering
       \labeledgraphics{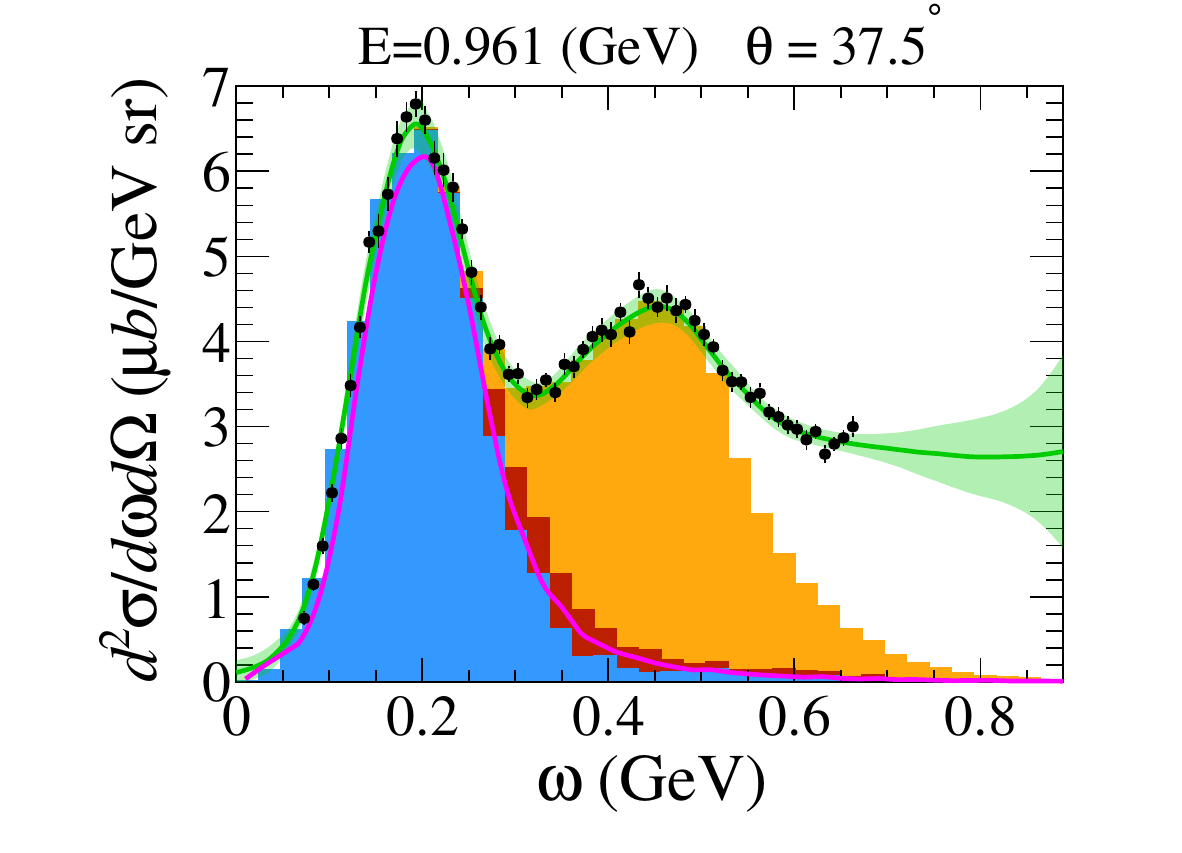}{0.900}{0.900}
       \label{fig:data_vs_dnn_vs_ewronet-a}
    \end{subfigure}
    \begin{subfigure}[b]{0.33\linewidth}
      \centering
      \labeledgraphics{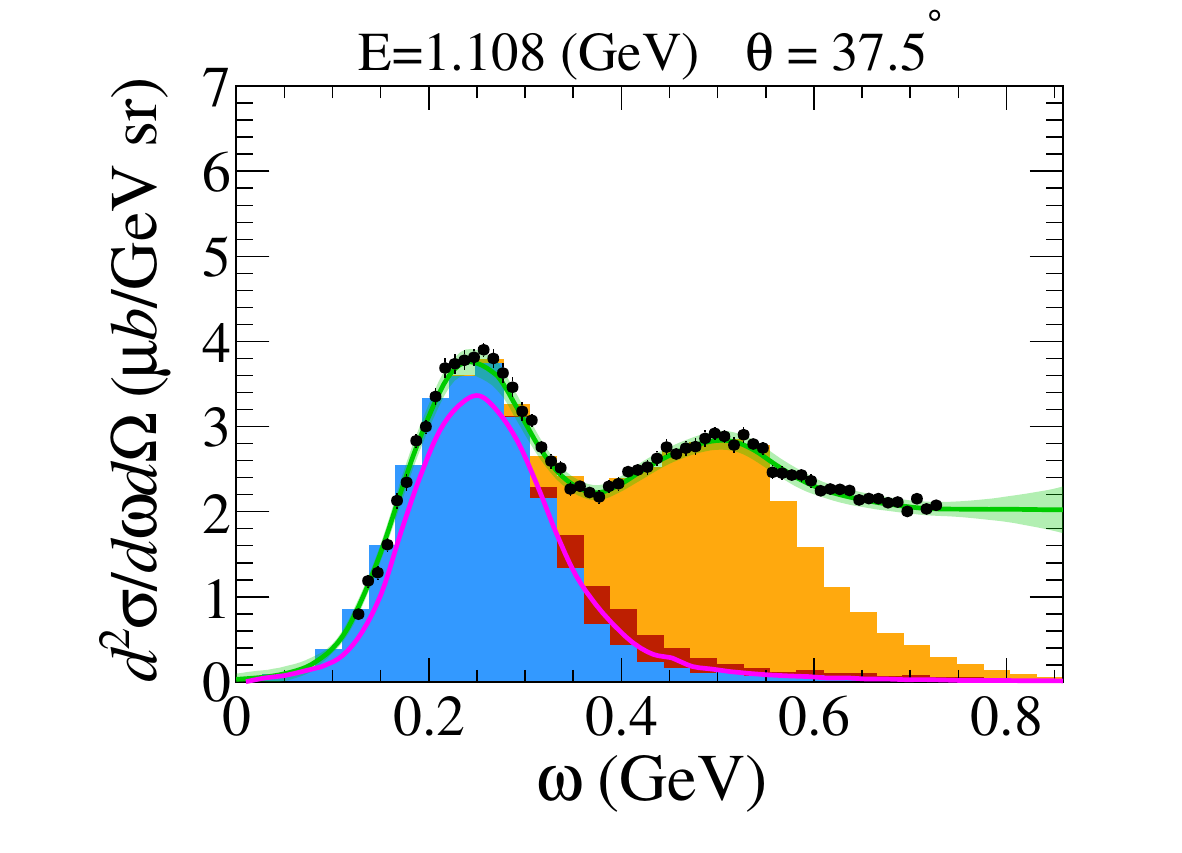}{0.900}{0.900}
      \label{fig:data_vs_dnn_vs_ewronet-b}
    \end{subfigure}
    \begin{subfigure}[b]{0.33\linewidth}
      \centering
       \citedlabeledgraphics{Figures/C12_E=1.299_theta=37.500.pdf}{0.598}{0.824}{Sealock:1989nx}{0.900}{0.900}
       \label{fig:data_vs_dnn_vs_ewronet-c}
    \end{subfigure}
  \caption{\label{fig:data_vs_dnn_vs_ewronet} (Color online) Double-differential cross section $d^{2}\sigma/d\omega\,d\Omega$ for inclusive electron scattering on carbon at $\theta=37.5^\circ$ for beam energies (a) $0.961$, (b) $1.108$, and (c) $1.299$~GeV. The stacked histograms show the \ewronet prediction decomposed into the \qe (blue), \mec (red), and \spp (orange) contributions; the magenta line shows the \qe SF prediction of \ewro~\cite{ankowski2025spectralfunctionapproachnuwro}; the green line and band show the DNN prediction of Ref.~\cite{Phys.Rev.C.112.055504} and its $1\sigma$ uncertainty. The data are from Sealock~\textit{et al.}~\cite{Sealock:1989nx}.}
  \end{figure*}

We begin with the inclusive cross sections for electron scattering on carbon measured at SLAC by Sealock~\textit{et al.}~\cite{Sealock:1989nx} at the scattering angle $\theta=37.5^\circ$ and beam energies $E=0.961$, $1.108$, and $1.299$~GeV. Figures~\ref{fig:data_vs_dnn_vs_ewronet-a}--\ref{fig:data_vs_dnn_vs_ewronet-c} show the double-differential cross section predicted by \ewronet, decomposed into the \qe, \mec, and \spp contributions, together with the DNN prediction (green line and band). For all three beam energies, the sum of the three channels reproduces the DNN prediction, within the statistical fluctuations of the MC sample, up to the upper boundary $\omega_\maxx$ of Eq.~(\ref{eq:omegamax}), which takes the values $\omega_\maxx=0.51$, $0.55$, and $0.61$~GeV, respectively. Since the DNN describes the data, \ewronet precisely reproduces the experimental data in this range as well. Beyond $\omega_\maxx$, the \ewronet prediction falls below the DNN, as expected from the design of the present scheme: the \spp model of \ewro includes only the $\Delta(1232)$ resonance and the nonresonant background, whereas at higher energy transfers the cross section receives increasing contributions from higher resonances, multipion production, and, eventually, \dis. Including these processes in \ewro would extend the kinematic region covered by \ewronet.

Figure~\ref{fig:data_vs_dnn_vs_ewronet} also shows how \ewronet distributes the inclusive strength among the exclusive channels. The \qe peak is described entirely by the \qe interactions, the \spp interactions account for most of the strength of the $\Delta(1232)$ peak, and the \mec contribution starts after the \qe peak, is concentrated in the dip region, and extends into the $\Delta(1232)$ region. The absence of \mec strength below the \qe peak is a consequence of our approximate \mec model, in which the onset of \mec is placed at the \qe peak by assuming that the peak originates purely from \qe scattering. For comparison, the figure also shows the \qe SF prediction of \ewro before reweighting (magenta line), which includes recent improvements~\cite{ankowski2025spectralfunctionapproachnuwro}. \qe SF reproduces the position of the \qe peak but lies below its measured height, leaving room for non-\qe contributions including 1-body and 2-body current interference at the peak. As noted in Ref.~\cite{ankowski2025spectralfunctionapproachnuwro}, non-\qe mechanisms contribute significantly at these kinematics, which suggests that the \mec contribution may set in below the \qe peak. A microscopic \mec model is therefore needed for a more reliable decomposition of the inclusive cross section.

\subsection{\label{subsection:ewronet-vs-clas} Comparison with the \clas and \efnu measurements}

\begin{figure}[htp!]
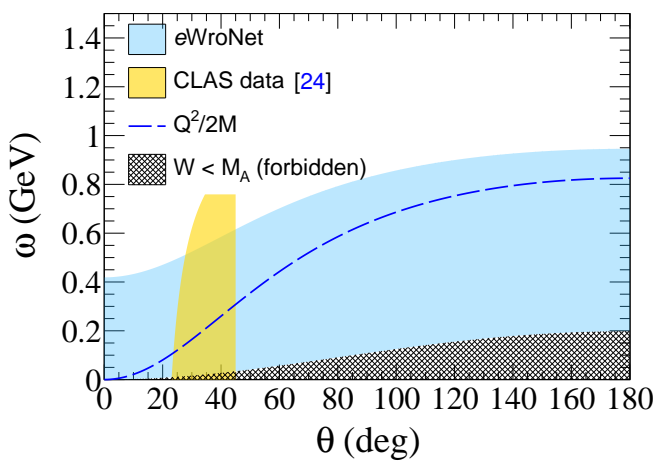

  \centering
  \citedgraphics{Figures/C12_E=1.159_region.pdf}{0.44}{0.8}{Khachatryan2021}
  \caption{\label{fig:region-coverance} (Color online) Kinematic coverage of \ewronet in the $(\theta,\omega)$ plane for $E=1.159$~GeV. The blue area denotes the region in which \ewronet perfectly reproduces the DNN~\cite{Phys.Rev.C.110.025501, Phys.Rev.C.112.055504}. The yellow area denotes the region spanned by the \clas data~\cite{Khachatryan2021}. The blue dashed line shows the position of the \qe peak for a free nucleon at rest, $\omega=Q^2/2M$. The hatched area, $W<M_A$, is kinematically forbidden.}
\end{figure}
We first examine to what extent the kinematic region probed by the \clas \eezeropi data is covered by \ewronet. Figure~\ref{fig:region-coverance} shows, for $E=1.159$~GeV, the region of the $(\theta,\omega)$ plane where \ewronet reproduces the DNN prediction. The yellow region, representing the \clas data, is bounded by $15^\circ\le\theta\le45^\circ$; by the minimum energy of the scattered electron, $E'\ge0.4$~GeV; and by momentum-dependent scattered-electron angle selection cuts from Eq.~(12) of Ref.~\cite{Khachatryan2021}, which further restrict the data to $\theta\gtrsim23^\circ$. The hatched area, where the invariant mass of the hadronic final state is below the target mass ($W<M_A$), is kinematically forbidden. The blue dashed line indicates the approximate position of the \qe peak, around which \qe scattering dominates. \ewronet covers the full angular range, with $\omega_\maxx$ increasing from $0.42$~GeV in the forward direction to $0.95$~GeV at backward angles, and therefore largely overlaps with the \clas data. The overlap includes the entire \qe region, the dip region, and the $\Delta(1232)$ region up to the right half-maximum of the peak. The part of the \clas phase space above $\omega_\maxx$, which extends up to $\omega=0.76$~GeV, is not covered by the present scheme, as it corresponds to the region of higher resonances. 

\begin{figure}[htp!]
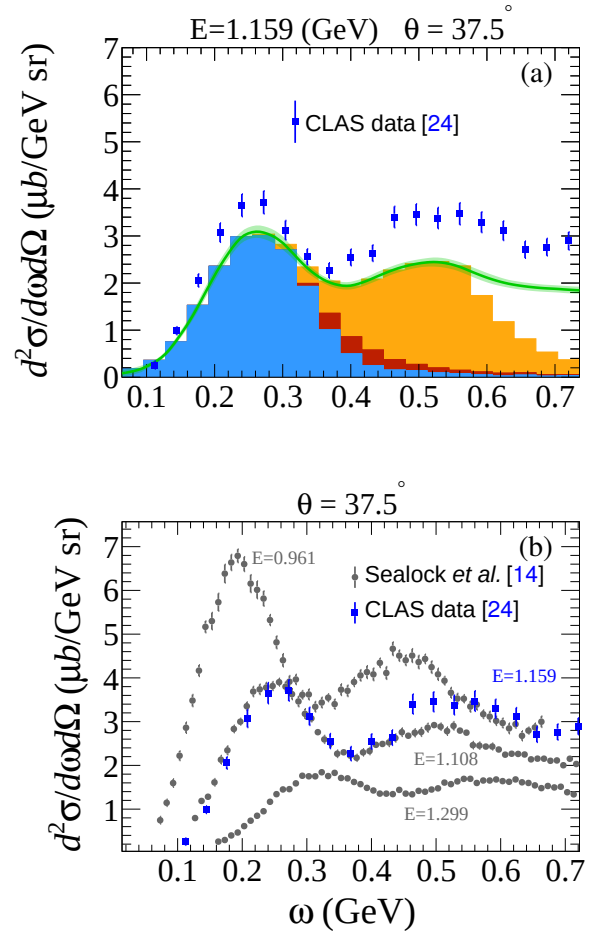

      \centering
      \begin{subfigure}[b]{\linewidth}
      \citedlabeledgraphics{Figures/C12_E=1.159_theta=37.500.pdf}{0.657}{0.722}{Khachatryan2021}{0.900}{0.900}
      \label{fig:data_vs_dnn_vs_ewronet-pt2-a}
      \end{subfigure}
      \begin{subfigure}[b]{\linewidth}
      \citedlabeledgraphics{Figures/ewronet_vs_slac_theta=37.5.pdf}{0.787}{0.771}{Sealock:1989nx}{0.749}{0.696}{Khachatryan2021}{0.900}{0.900}
      \label{fig:data_vs_dnn_vs_ewronet-pt2-b}
      \end{subfigure}
    \caption{\label{fig:data_vs_dnn_vs_ewronet-pt2} (Color online) (a) Same as Fig.~\ref{fig:data_vs_dnn_vs_ewronet}, but for $E=1.159$~GeV and $\theta=37.5^\circ$, compared with the inclusive cross section measured by the \clas and \efnu collaborations~\cite{Khachatryan2021}. (b) Overlay of the data points of Sealock~\textit{et al.}~\cite{Sealock:1989nx} and of the \clas and \efnu collaborations~\cite{Khachatryan2021}.}
\end{figure}
We then compare \ewronet with the inclusive cross section measured by the \clas and \efnu collaborations at JLab~\cite{Khachatryan2021} for $E=1.159$~GeV and $\theta=37.5^\circ$, shown in Fig.~\ref{fig:data_vs_dnn_vs_ewronet-pt2-a}. While \ewronet reproduces the DNN prediction up to $\omega_\maxx$, the measured cross section lies above both. The data agree with the predictions on the rising edge of the \qe peak but depart from them quickly at higher energy transfers: both \ewronet and the DNN lie about $15\%$ below the measurement at the \qe peak and about $30\%$ below it at the $\Delta(1232)$ peak, well outside both the DNN uncertainty and the normalization uncertainty of the data, $\approx3\%$~\cite{Khachatryan2021}.
One should view this difference in light of the comparison with the data of Sealock~\textit{et al.}~\cite{Sealock:1989nx}. In Fig.~\ref{fig:data_vs_dnn_vs_ewronet-pt2-b}, the kinematics of the \clas measurement lie between the settings $E=1.108$ and $1.299$~GeV of Sealock~\textit{et al.}~\cite{Sealock:1989nx} at the same scattering angle, both of which are reproduced perfectly by the DNN as well as \ewronet (see Fig.~\ref{fig:data_vs_dnn_vs_ewronet}).  The DNN prediction at $1.159$~GeV should therefore be viewed as an interpolation between the kinematics $E=1.108$ and $1.299$~GeV. As $E$ increases, the peaks from \qe and \spp interactions decrease, while their positions shift toward higher energy transfer. However, at $E=1.159$~GeV, the \qe peak height is similar to that at $E=1.108$~GeV, while the $\Delta(1232)$ peak strength is significantly higher. The difference in the trend suggests tension between the \clas measurement~\cite{Khachatryan2021} and the data from Sealock~\textit{et al.}~\cite{Sealock:1989nx} at neighboring kinematics, the origin of which is beyond the scope of the present work.

\begin{figure}[htp!]
    \centering
        \begin{subfigure}[b]{\linewidth}
       \citedlabeledgraphics{Figures/C12_E=1.159_Erec.pdf}{0.436}{0.712}{Khachatryan2021}{0.905}{0.849}
       \label{fig:CLAS-Erec-a}
    \end{subfigure}
    \hfill
    \begin{subfigure}[b]{\linewidth}
       \labeledgraphics{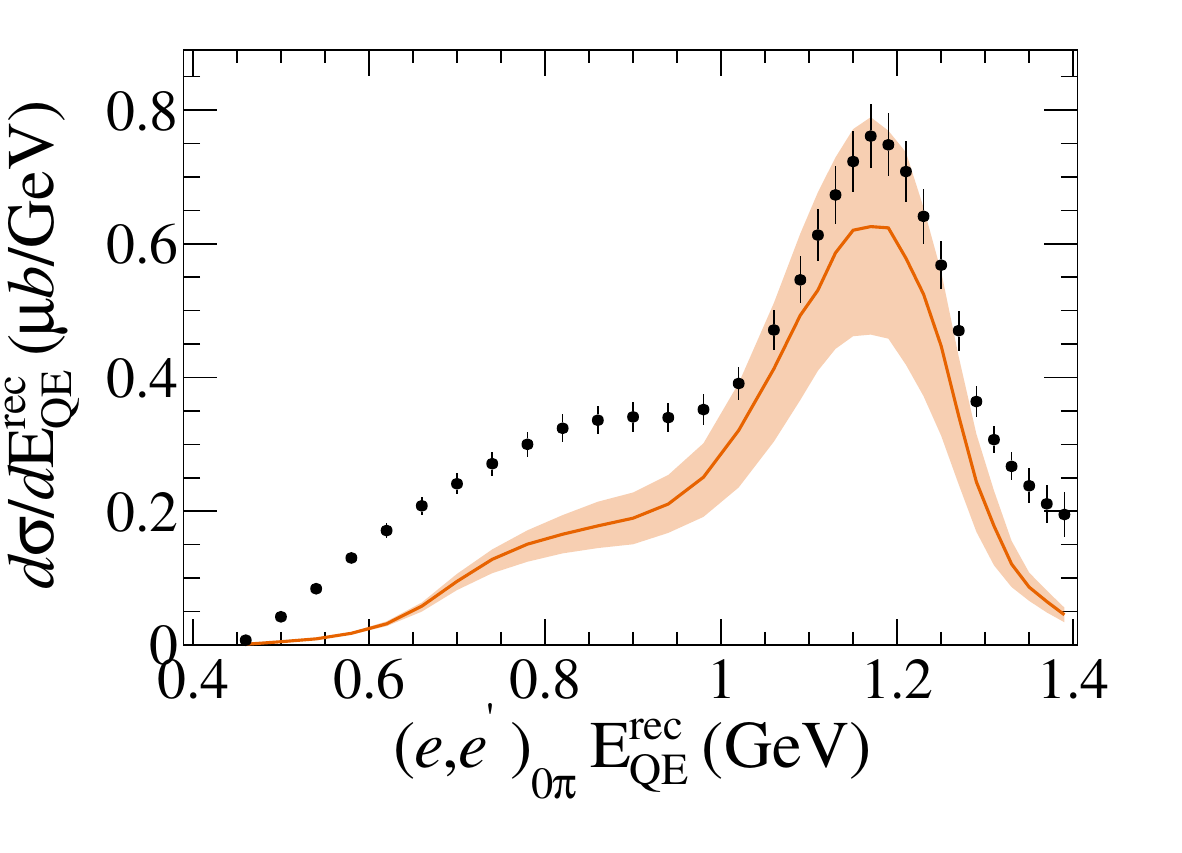}{0.913}{0.943}
       \label{fig:CLAS-Erec-b}
    \end{subfigure}
    \caption{\label{fig:CLAS-Erec} (Color online) Distribution of the reconstructed energy $\qeerec$, Eq.~(\ref{eqn:Erec}), for the $(e,e')_{0\pi}$ sample on carbon at $E=1.159$~GeV and $15^\circ\le\theta\le45^\circ$, compared with the data of the \clas and \efnu collaborations~\cite{Khachatryan2021}. (a) \ewronet prediction decomposed into the \qe (blue), \mec (red), and \spp (orange) contributions. (b) Total \ewronet prediction with the band obtained by shifting the DNN prediction by $\pm1\sigma$.}
\end{figure}
Finally, we compare the \ewronet predictions with the semi-exclusive measurement of the \clas and \efnu collaborations~\cite{Khachatryan2021}. Figure~\ref{fig:CLAS-Erec} shows the $\qeerec$ distribution for the \eezeropi sample. This comparison highlights a key feature of the scheme: through \ewronet, the inclusive DNN model can be confronted with semi-exclusive data. As seen in Fig.~\ref{fig:CLAS-Erec-a}, the \qe interactions produce a peak centered near the beam energy, broadened by the Fermi motion of the nucleons, and its low-energy side is populated by events with large removal energies, particularly those involving short-range correlations. At lower energies, the tail of the $\qeerec$ distribution is populated mainly by non-\qe events that pass the pionless selection. The \mec interactions contribute mainly just below the \qe peak, whereas the \spp interactions, in which the pion is absorbed in the nucleus or falls below the detection threshold, dominate from $\qeerec\lesssim0.9$~GeV. 

The \ewronet prediction falls short of the data by about $20\%$ at the \qe peak and by about $45\%$ in the low-$\qeerec$ tail, exceeding the deficit found in the inclusive comparison. Several effects may contribute to this larger difference. For both the \qe- and \spp-dominated regions, the semi-exclusive distribution integrates over the full angular range $15^\circ\le\theta\le45^\circ$ rather than a single scattering angle. For \spp, the contribution to the pionless sample depends not only on the inclusive strength but also on the description of pion absorption in the intranuclear cascade. In addition, the region $\qeerec\lesssim0.7$~GeV also receives contributions from higher resonances and multipion production at $\omega>\omega_\maxx$, which lie outside the kinematic region of \ewronet. The overall deficit, however, is largely independent of possible mismodeling of \mec, since a different \mec model would redistribute the inclusive strength among the exclusive channels and alter the kinematics of the knocked-out nucleons, but not the total strength, which is fixed by the DNN.

The DNN~\cite{Phys.Rev.C.110.025501,Phys.Rev.C.112.055504} also provides the uncertainty of its prediction, Eq.~(\ref{eqn:uncertainty-dnn-curve}), which can be used to shift the inclusive cross section by $\pm1\sigma$. Figure~\ref{fig:CLAS-Erec-b} shows the total \ewronet prediction, summed over all channels, for the mean DNN prediction, together with the band obtained by shifting the DNN prediction by $\pm1\sigma$. At forward scattering angles, where DNN uncertainties are larger, the \qe contribution is much larger than the \spp one, so shifting the inclusive cross section there affects \qe interactions more. Within this uncertainty, \ewronet reproduces the \qe peak region. This is not the case in the $\Delta(1232)$ region, where the band is narrower, and the prediction still lies about $40\%$ below the measurement. Overall, the semi-exclusive comparison reflects a normalization difference relative to the \clas data similar to the inclusive one and points to modeling pion absorption and the region beyond $\omega_\maxx$ as the main directions for improvement.

\section{\label{sec:discussion}DISCUSSION}

The DNN models have so far been used as stand-alone predictions of the  inclusive electron-scattering cross section, compared either with data or with theoretical calculations~\cite{Phys.Rev.C.110.025501, Phys.Rev.C.112.055504, Phys.Lett.B.859.139142, Phys.Rev.Lett.052502}. In event generation, machine learning has entered mostly through generative models trained on the output of existing generators or theoretical models, reproducing the physics they contain at lower computational cost~\cite{Bonilla2025GAN, Bonilla2026TL, ElBaz2025}.

In this article, we proposed a scheme to decompose the inclusive DNN model~\cite{Phys.Rev.C.110.025501,Phys.Rev.C.112.055504} into exclusive reaction channels and use this decomposition to reweight the theoretical models of the \ewro MC generator, which provides nuclear modeling and kinematics of final-state particles. We tested the scheme against inclusive measurements to demonstrate its capabilities, and then compared \ewronet with the measurements from the \clas and \efnu collaborations~\cite{Khachatryan2021}, motivated by the wide attention received by its conclusions on neutrino energy reconstruction. For the semi-exclusive observables, the comparison is limited by the kinematic coverage of the scheme, which does not extend beyond $\omega_\maxx$. Within the DNN uncertainties, which are large at forward scattering angles where measurements are scarce~\cite{Phys.Rev.C.112.055504}, \ewronet reproduces the \qe peak in the $\qeerec$ distribution, whereas it still underestimates the $\Delta(1232)$ region, partially because of the missing coverage. The overall picture is not yet fully consistent, but \ewronet makes it possible to compare DNN predictions with semi-exclusive data. The approximate treatment of \mec remains an inherent weakness of the present model, which makes it difficult to draw definite conclusions.

We are aware that the decomposition proposed here has inherent limitations. The interaction channels are not strictly separable, because the amplitudes of different mechanisms leading to the same final state interfere quantum mechanically; the contributions of one- and two-nucleon currents, for instance, have to be treated consistently at the amplitude level~\cite{Phys.Rev.Lett.116.192501}. \qe scattering leads not only to one-nucleon but also to two-nucleon knockout because of short-range correlations in the target nucleus~\cite{ankowski2025spectralfunctionapproachnuwro}, while \mec can emit two or more nucleons~\cite{Martini:2009uj}. These effects are difficult to address in the present scheme, which treats the channels as incoherent contributions, and we regard them as a limitation of the approach.

The present implementation uses the DNN ensemble for the carbon target. Networks for other nuclei obtained with transfer learning~\cite{Phys.Rev.Lett.052502} can be included in the same way. An argon model, in which the inclusive cross section is given by a DNN and the nuclear model by the argon SF~\cite{ankowski2025spectralfunctionapproachnuwro}, would be of immediate interest for \dune. Extending beyond $\omega_\maxx$ requires dedicated channels for higher resonances and \dis in \ewro.

\begin{acknowledgments}
This work was partly (K.M.G., B.E.K., and J.T.S.) or fully (R.D.B., J.L.B., and
H.P.) supported by the National Science Centre, Poland, under
grant UMO-2021/41/B/ST2/02778.
\end{acknowledgments}

\appendix

\section{\label{sec:MEC} Approximate model of \mec}

\begin{figure}[htbp!]
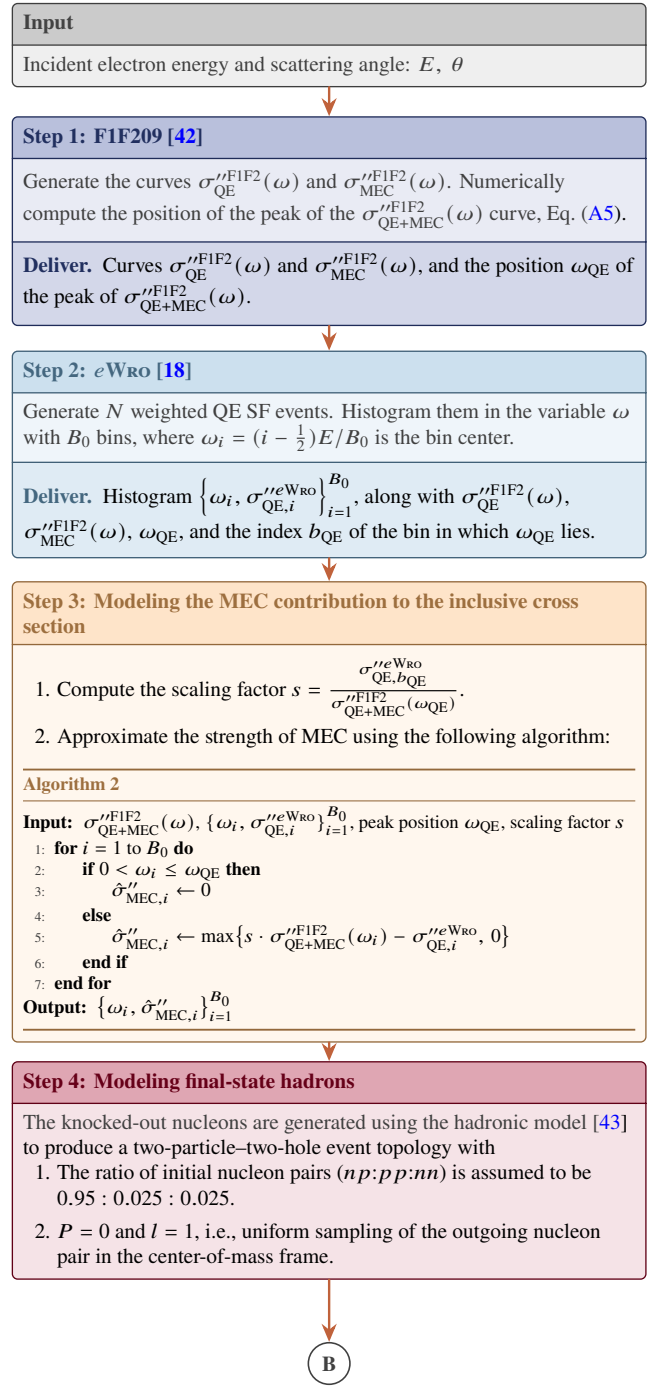

    \centering
    \begin{blockdiagram}

        \bdbox[color=sunGrey, width=8.1cm, anchor=north]{minput}{0,0}{Input}
              {Incident electron energy and scattering angle: $E,\,\theta$}

        \bdbox[color=sunIndigo, width=8.1cm, anchor=north,
               deliver={Curves $\xF{\QE}(\omega)$ and $\xF{\MEC}(\omega)$, and the
                        position $\omega_\QE$ of the peak of $\xF{\QE+\MEC}(\omega)$.}]%
              {mf1f2}{0,-1.5}{Step~1: F1F209~\cite{bosted}}%
              {Generate the curves $\xF{\QE}(\omega)$ and $\xF{\MEC}(\omega)$.
               Numerically compute the position of the peak of the
               $\xF{\QE+\MEC}(\omega)$ curve, Eq.~(\ref{eq:qelike}).}

        \bdbox[color=sunSky, width=8.1cm, anchor=north,
               deliver={Histogram $\left\{\omega_i,\xE{\QE,i}\right\}_{i=1}^{B_0}$,
                        along with $\xF{\QE}(\omega)$, $\xF{\MEC}(\omega)$, $\omega_\QE$,
                        and the index $b_\QE$ of the bin in which $\omega_\QE$ lies.}]%
              {mew}{0,-4.6}{Step~2: \ewro~\cite{NuWro_GitHub_2025}}%
              {Generate $N$ weighted \qe SF events. Histogram them in the variable
               $\omega$ with $B_0$ bins, where $\omega_i=(i-\frac{1}{2})E/B_0$ is the
               bin center.}

        \bdbox[color=sunAmber, width=8.1cm, anchor=north]%
              {mmec}{0,-7.65}{Step~3: Modeling the \mec contribution to the inclusive cross section}%
              {

              \begin{bdenumerate}
                \item Compute the scaling factor $s = \frac{\xE{\QE,b_\QE}}{\xF{\QE+\MEC}(\omega_\QE)}$.
                \item Approximate the strength of \mec using the following algorithm:
              \end{bdenumerate}
                \begin{bdalgo}[color=sunAmber, label=alg:approx-mec]{}
                 \Require $\xF{\QE+\MEC}(\omega)$, $\{\omega_i,\xE{\QE,i}\}_{i=1}^{B_0}$,
                          peak position $\omega_{\QE}$, scaling factor $s$
                 \For{$i = 1$ to $B_0$}
                   \If{$0 < \omega_i \le \omega_{\QE}$}
                     \State $\xT{\MEC,i} \gets 0$
                   \Else
                     \State $\xT{\MEC,i} \gets
                            \max\bigl\{s\cdot\xF{\QE+\MEC}(\omega_i)-\xE{\QE,i},\,0\bigr\}$
                   \EndIf
                 \EndFor
                 \Ensure $\bigl\{\omega_i,\xT{\MEC,i}\bigr\}_{i=1}^{B_0}$
               \end{bdalgo}

              }

        \bdbox[color=sunMaroon, width=8.1cm, anchor=north]{mhad}{0,-14}%
              {Step~4: Modeling final-state hadrons}%
              {The knocked-out nucleons are generated using the hadronic model~\cite{PhysRevD.111.036032} to produce a two-particle--two-hole event topology with
              \begin{bdenumerate}
                \item The ratio of initial nucleon pairs ($np$:$pp$:$nn$) is assumed to be $0.95:0.025:0.025$.
                \item $P=0$ and $l=1$, i.e., uniform sampling of the outgoing nucleon pair in the center-of-mass frame.
              \end{bdenumerate}}

        \bdarrow{minput}{mf1f2}
        \bdarrow{mf1f2}{mew}
        \bdarrow{mew}{mmec}
        \bdarrow{mmec}{mhad}

        \bdconnector{connB}{0,-18}{B}
        \bdarrow{mhad}{connB}

    \end{blockdiagram}
    \caption{\label{fig:mec-block-diagram} (a) Block diagram of the approximate \mec model in \ewronet, part one: inputs, \ewro, approximation of the \mec strength using F1F209~\cite{bosted}, and production of final-state particles using the hadronic model~\cite{PhysRevD.111.036032}. The flow continues at connector~B. }
\end{figure}

\begin{figure}[!t]
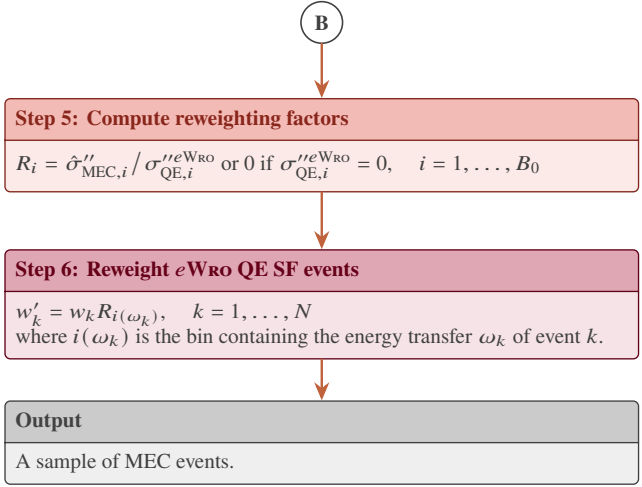

  \ContinuedFloat
  \begin{blockdiagram}
        \bdbox[color=sunRed, width=8.1cm, anchor=north]{mrf}{0,-1.}%
              {Step~5: Compute reweighting factors}%
              {$R_{i}=\xT{\MEC, i}\big/\xE{\QE,i}$ or 0 if $\xE{\QE, i} = 0$, \quad
               $i=1,\dots,B_{0}$}

        \bdbox[color=sunMaroon, width=8.1cm, anchor=north]{rw}{0,-3.0}%
              {Step~6: Reweight \ewro \qe SF events}%
              {$w'_{k}=w_{k}R_{i(\omega_{k})}$, \quad $k=1,\dots,N$\\
               where $i(\omega_k)$ is the bin containing
               the energy transfer $\omega_k$ of event $k$.}

        \bdbox[color=sunGrey, width=8.1cm, anchor=north]{mout}{0,-5.0}%
              {Output}%
              {A sample of \mec events.}

        \bdconnector{connB}{0,0}{B}
        \bdarrow{connB}{mrf}
        \bdarrow{mrf}{rw}
        \bdarrow{rw}{mout}

    \end{blockdiagram}
    \caption{\label{fig:mec-block-diagram-pt2}(Continued.) (b) Part two: the reweighting factors (Step~5) and the reweighting of the \ewro \qe SF events (Step~6), which yield the sample of \mec events, starting from connector~B of part (a).}
\end{figure}

We construct an approximate \mec model that estimates the contribution of \mec to the inclusive cross section. The starting point is the empirical parametrization of inclusive electron-nucleus scattering developed by Bosted and Mamyan~\cite{bosted}, referred to as F1F209. The fit was obtained from the world data on nuclei with $A>2$ and is valid for $0<W<3.2$~GeV and $0.2<Q^{2}<5$~GeV$^{2}$, where $W^{2}=M^{2}+2M\omega-Q^{2}$ is the squared invariant mass of the hadronic system and $M$ is the average nucleon mass. The cross section is expressed through the nuclear structure functions $W_{1}$ and $W_{2}$ as:

\begin{equation}
  \frac{d^{2}\sigma}{d\omega\, d\Omega}
  = \sigma_{\rm Mott}\left[\,W_{2}(W^{2},Q^{2})
  + 2\tan^{2}(\theta/2)\,W_{1}(W^{2},Q^{2})\right],
  \label{eq:rosenbluth}
\end{equation}
where $\sigma_{\rm Mott}$ is the Mott cross section. Each structure function is the sum of a \qe and an inelastic part, $W_{1,2}=W_{1,2}^{\rm QE}+W_{1,2}^{I}$.

The key feature that this study exploits is that the inelastic part of $W_{1}$ in the fit is constructed as
\begin{equation}
  W_{1}^{I} = \left[\,W_{1}^{\rm F}(W^{2},Q^{2})
  + W_{1}^{\rm MEC}(W^{2},Q^{2})\right] f_{\rm EMC}(x'),
  \label{eq:w1inel}
\end{equation}
while $W_{2}^{I}$ is obtained from it as
\begin{equation}
    W_{2}^{I} = W_{1}^{I}\,\frac{1+R_{A}(W,Q^{2})}{1+\omega^{2}/Q^{2}}.
\end{equation}
Here, $W_{1}^{\rm F}$ is a Fermi-smeared incoherent sum of the free proton and neutron structure functions, $W_{1}^{\rm MEC}$ is an additional term introduced to fill the dip between the \qe and $\Delta(1232)$ peaks, $f_{\rm EMC}$ parametrizes the nuclear modification of the \dis structure functions, and $R_{A}(W,Q^2)$ is the nuclear longitudinal-to-transverse cross section ratio. As $W_{1}^{\rm MEC}$ enters Eq.~(\ref{eq:w1inel}) additively and the factor converting $W_{1}^{I}$ into $W_{2}^{I}$ is common to both terms, the inelastic cross section can be separated as
\begin{equation}
  \frac{d^{2}\sigma^{I}}{d\omega\,d\Omega}
  = \frac{d^{2}\sigma^{\INEL }}{d\omega\,d\Omega}
  + \frac{d^{2}\sigma^{\rm MEC}}{d\omega\,d\Omega},
  \label{eq:split}
\end{equation}
where \inel collects the $\Delta(1232)$ region, the higher resonances, and the \dis continuum.

Together with the \qe part, the fit thus delivers three curves at fixed $(E,\theta)$:
\[
  \xF{\QE}(\omega),\, \xF{\MEC}(\omega),\, \xF{\INEL}(\omega)
\]
of which only $\xF{\QE}$ and $\xF{\MEC}$ are used in this work. We now proceed as follows:

In Step~1, F1F209 provides the curves $\xF{\QE}(\omega)$ and $\xF{\MEC}(\omega)$ at the chosen $(E,\theta)$, from which we form the QE-like strength
\begin{equation}
  \xF{\QE+\MEC}(\omega) \equiv \xF{\QE}(\omega) + \xF{\MEC}(\omega)
  \label{eq:qelike}
\end{equation}
We then determine numerically the peak position $\omega_\QE$ of $\xF{\QE+\MEC}(\omega)$. In Step~2, \ewro generates $N$ weighted \qe events within the SF approach (Sect.~\ref{subsubsec:QESF}). The events are histogrammed in $\omega$ in $B_0$ bins to give $\{\omega_i, \xE{\QE,i}\}_{i=1}^{B_0}$, where $\omega_i$ denotes the bin center of the $i$th bin and $\xE{\QE, i}$ is the corresponding \qe SF cross section. The bin containing $\omega_\QE$ is denoted by $b_\QE$.

In Step~3, the curve $\xF{\QE+\MEC}$ is scaled by a factor $s$:
\begin{equation}
  s=\xE{\QE,b_\QE}/\xF{\QE+\MEC}(\omega_\QE),
\end{equation}
so that the scaled curve matches the \qe SF cross section at the peak position $\omega_\QE$. Below $\omega_\QE$, the \mec contribution is set to zero. Above $\omega_\QE$, the remaining normalization is attributed to \mec:
\begin{equation}
  \xT{\MEC,i}=\max\{s\,\xF{\QE+\MEC}(\omega_i)-\xE{\QE,i},\,0\}
\end{equation}
In Step~4, the lepton kinematics of each \qe event are kept, while two outgoing nucleons are generated using the hadronic model of Ref.~\cite{PhysRevD.111.036032}, assuming $np$, $pp$, and $nn$ initial pairs in the proportion $0.95:0.025:0.025$ and no nucleon-nucleon correlations, i.e., isotropic emission in the center-of-mass frame of the pair. Finally, in Steps~5 and 6, the events are reweighted with $R_i=\xT{\MEC,i}/\xE{\QE,i}$ and flagged as \mec, such that the generated sample reproduces $\xT{\MEC,i}$.

\section{\label{sec:delta-peak-width-derivation} Position and width of the \texorpdfstring{$\Delta(1232)$}{Delta(1232)} peak}

We consider the reaction
\begin{equation}
  e(k) + N(p) \rightarrow e'(k') + X(p_X),
  \label{eqn:reaction}
\end{equation}
in which an electron with four-momentum $k=(E,\mathbf{k})$ scatters off a nucleon $N$ bound in the nucleus and emerges with $k'=(E',\mathbf{k}')$ at the angle $\theta$, while $X$ denotes the remaining hadronic final state. The four-momentum transfer is $q=k-k'=(\omega,\mathbf{q})$, with $\omega=E-E'$, and, neglecting the electron mass,
\begin{equation}
  Q^2 \equiv -q^2 = 4EE'\sin^2(\theta/2).
\end{equation}
To locate the $\Delta(1232)$ peak, we neglect nuclear binding and Fermi motion, i.e., we write the invariant mass of the hadronic system, $W^2=(p+q)^2$, assuming that the scattering takes place on a free nucleon of mass $M$ at rest, $p=(M,\mathbf{0})$. Using $E'=E-\omega$, one finds
\begin{equation}
  \begin{split}
  W^2 &= M^2 + 2M\omega - Q^2 \\
      &= M^2 - 4E^2\sin^2(\theta/2) + \left(2M + 4E\sin^2(\theta/2)\right)\omega,
  \end{split}
  \label{eqn:W2-function-w}
\end{equation}
so that, at fixed $(E,\theta)$, $W^2$ is a linear function of $\omega$. Setting $W=M_\Delta$ gives the position of the peak,
\begin{equation}
  \omega_\Delta = \frac{M_\Delta^2 - M^2 + 4E^2\sin^2(\theta/2)}{2M + 4E\sin^2(\theta/2)}.
  \label{eqn:peakposition-Delta}
\end{equation}

To estimate the width of the peak, we approximate the response in the $\Delta$ region by a relativistic Breit--Wigner function with a constant width,
\begin{equation}
  \mathcal{B}(W^2) = \frac{K}{\left(W^2 - M_\Delta^2\right)^2 + M_\Delta^2\Gamma_0^2},
  \label{eqn:breit-wigner}
\end{equation}
where $K$ does not depend on $W$ and $\Gamma_0=0.12$~GeV is the free width of the $\Delta(1232)$. The function reaches its maximum, $\mathcal{B}_{\max}=K/(M_\Delta^2\Gamma_0^2)$, at $\omega=\omega_\Delta$. Because Eq.~(\ref{eqn:W2-function-w}) implies
\begin{equation}
  W^2 - M_\Delta^2 = \left(2M + 4E\sin^2(\theta/2)\right)(\omega - \omega_\Delta),
\end{equation}
the half-maximum condition, $\mathcal{B}=\mathcal{B}_{\max}/2$, or equivalently $|W^2-M_\Delta^2| = M_\Delta\Gamma_0$, is satisfied at
\begin{equation}
  \omega_\pm = \omega_\Delta \pm \frac{M_\Delta\Gamma_0}{2M + 4E\sin^2(\theta/2)}.
  \label{eqn:omega-pm}
\end{equation}
The right half-maximum point, $\omega_+$, defines the upper boundary $\omega_{\max}$ of Eq.~(\ref{eq:omegamax}), up to which the \ewro events are reweighted to reproduce the inclusive cross section of the DNN.

\bibliography{main.bib}

@misc{bosted,
      title={Empirical Fit to electron-nucleus scattering}, 
      author={P. E. Bosted and V. Mamyan},
      year={2012},
      eprint={1203.2262},
      archivePrefix={arXiv},
      primaryClass={nucl-th},
      url={https://arxiv.org/abs/1203.2262}, 
}

@article{Phys.Rev.C.110.025501,
  title = {Empirical fits to inclusive electron-carbon scattering data obtained by deep-learning methods},
  author = {Kowal, Beata E. and Graczyk, Krzysztof M. and Ankowski, Artur M. and Banerjee, Rwik Dharmapal and Prasad, Hemant and Sobczyk, Jan T.},
  journal = {Phys. Rev. C},
  volume = {110},
  issue = {2},
  pages = {025501},
  numpages = {11},
  year = {2024},
  month = {Aug},
  publisher = {American Physical Society},
  doi = {10.1103/PhysRevC.110.025501},
  url = {https://link.aps.org/doi/10.1103/PhysRevC.110.025501}
}

@article{Phys.Rev.C.112.055504,
  title = {Re-optimization of a deep neural network model for electron-carbon scattering using new experimental data},
  author = {Kowal, Beata E. and Graczyk, Krzysztof M. and Ankowski, Artur M. and Banerjee, Rwik Dharmapal and Bonilla, Jose L. and Prasad, Hemant and Sobczyk, Jan T.},
  journal = {Phys. Rev. C},
  volume = {112},
  issue = {5},
  pages = {055504},
  numpages = {14},
  year = {2025},
  month = {Nov},
  publisher = {American Physical Society},
  doi = {10.1103/ydvc-2567},
  url = {https://link.aps.org/doi/10.1103/ydvc-2567}
}

@article{Phys.Lett.B.859.139142,
title = {Modeling inclusive electron-nucleus scattering with Bayesian artificial neural networks},
journal = {Physics Letters B},
volume = {859},
pages = {139142},
year = {2024},
issn = {0370-2693},
doi = {https://doi.org/10.1016/j.physletb.2024.139142},
url = {https://www.sciencedirect.com/science/article/pii/S0370269324007007},
author = {Joanna E. Sobczyk and Noemi Rocco and Alessandro Lovato},
}

@article{_muda_2015,
   title={{\tt NuWro} Monte Carlo Generator of Neutrino Interactions --- First Electron Scattering Results},
   volume={46},
   ISSN={1509-5770},
   url={http://dx.doi.org/10.5506/APhysPolB.46.2329},
   DOI={10.5506/aphyspolb.46.2329},
   number={11},
   journal={Acta Physica Polonica B},
   publisher={Jagiellonian University},
   author={Żmuda, J. and Graczyk, K.M. and Juszczak, C. and Sobczyk, J.T.},
   year={2015},
   pages={2329} }

@article{Nucl.Phys.A.505.267,
title = {The nucleon spectral function in nuclear matter},
journal = {Nuclear Physics A},
volume = {505},
number = {2},
pages = {267-299},
year = {1989},
issn = {0375-9474},
doi = {https://doi.org/10.1016/0375-9474(89)90374-6},
url = {https://www.sciencedirect.com/science/article/pii/0375947489903746},
author = {Omar Benhar and Adelchi Fabrocini and Stefano Fantoni},
}

@article{Nucl.Phys.A.579.493,
title = {Spectral function of finite nuclei and scattering of {GeV} electrons},
journal = {Nuclear Physics A},
volume = {579},
number = {3},
pages = {493-517},
year = {1994},
issn = {0375-9474},
doi = {https://doi.org/10.1016/0375-9474(94)90920-2},
url = {https://www.sciencedirect.com/science/article/pii/0375947494909202},
author = {O. Benhar and A. Fabrocini and S. Fantoni and I. Sick},
}

@article{Phys.Rev.Lett.052502,
  title = {Electron-Nucleus Cross Sections from Transfer Learning},
  author = {Graczyk, Krzysztof M. and Kowal, Beata E. and Ankowski, Artur M. and Banerjee, Rwik Dharmapal and Bonilla, Jose Luis and Prasad, Hemant and Sobczyk, Jan T.},
  journal = {Phys. Rev. Lett.},
  volume = {135},
  issue = {5},
  pages = {052502},
  numpages = {5},
  year = {2025},
  month = {Aug},
  publisher = {American Physical Society},
  doi = {10.1103/zxv6-22tz},
  url = {https://link.aps.org/doi/10.1103/zxv6-22tz}
}

@misc{NuWro_GitHub_2025,
  author       = {Sobczyk, Jan},
  title        = {{NuWro: A Monte Carlo Neutrino Event Generator}},
  year         = {2025},
  url          = {https://github.com/NuWro/nuwro},
  note         = {GitHub repository. Accessed: 2026-01-22},
  version      = {latest commit at time of access},
  organization = {University of Wroc{\l}aw / NuWro Contributors}
}

@misc{ankowski2025spectralfunctionapproachnuwro,
      title={Spectral function approach in {NuWro}: modeling of multinucleon final states in quasielastic scattering}, 
      author={Artur M. Ankowski and Rwik Dharmapal Banerjee and Jan T. Sobczyk and José L. Bonilla and Krzysztof M. Graczyk and Beata E. Kowal and Hemant Prasad},
      year={2025},
      eprint={2508.10101},
      archivePrefix={arXiv},
      primaryClass={hep-ph},
      url={https://arxiv.org/abs/2508.10101}, 
}

@article{PhysRevD.109.073004,
  title = {JLab spectral functions of argon in {NuWro} and their implications for {MicroBooNE}},
  author = {Banerjee, Rwik Dharmapal and Ankowski, Artur M. and Graczyk, Krzysztof M. and Kowal, Beata E. and Prasad, Hemant and Sobczyk, Jan T.},
  journal = {Phys. Rev. D},
  volume = {109},
  issue = {7},
  pages = {073004},
  numpages = {10},
  year = {2024},
  month = {Apr},
  publisher = {American Physical Society},
  doi = {10.1103/PhysRevD.109.073004},
  url = {https://link.aps.org/doi/10.1103/PhysRevD.109.073004}}

@article{Bonilla2025GAN,
  title = {Generative adversarial neural networks for simulating neutrino interactions},
  author = {Bonilla, Jos\'e L. and Graczyk, Krzysztof M. and Ankowski, Artur M. and Banerjee, Rwik Dharmapal and Kowal, Beata E. and Prasad, Hemant and Sobczyk, Jan T.},
  journal = {Phys. Rev. D},
  volume = {112},
  issue = {1},
  pages = {013007},
  year = {2025},
  publisher = {American Physical Society},
  doi = {10.1103/l6td-93sr},
  url = {https://link.aps.org/doi/10.1103/l6td-93sr}
}

@article{Bonilla2026TL,
  title = {Transfer learning for neutrino scattering: Domain adaptation with generative adversarial networks},
  author = {Bonilla, Jos\'e L. and Graczyk, Krzysztof M. and Ankowski, Artur M. and Banerjee, Rwik Dharmapal and Kowal, Beata E. and Prasad, Hemant and Sobczyk, Jan T.},
  journal = {Phys. Rev. D},
  volume = {113},
  issue = {5},
  pages = {053001},
  year = {2026},
  publisher = {American Physical Society},
  doi = {10.1103/kwjj-wp1c},
  url = {https://link.aps.org/doi/10.1103/kwjj-wp1c}
}

@article{ElBaz2025,
  title = {Efficient {Monte Carlo} event generation for neutrino-nucleus exclusive cross sections},
  author = {El Baz, Mathias and S\'anchez, Federico and Jachowicz, Natalie and Niewczas, Kajetan and Jha, Ashish Kumar and Nikolakopoulos, Alexis},
  journal = {Phys. Rev. D},
  volume = {111},
  issue = {11},
  pages = {113001},
  year = {2025},
  publisher = {American Physical Society},
  doi = {10.1103/v7sz-vn3b},
  url = {https://link.aps.org/doi/10.1103/v7sz-vn3b}
}

@misc{onnxruntime,
  author = {{ONNX Runtime developers}},
  title = {{ONNX Runtime}},
  year = {2021},
  howpublished = {\url{https://onnxruntime.ai/}}
}

@article{Khachatryan2021,
  title         = {Electron-beam energy reconstruction for neutrino oscillation measurements},
  author        = {Khachatryan, M. and Papadopoulou, A. and Ashkenazi, A. and Hauenstein, F. and Nambrath, A. and Hrnjic, A. and Weinstein, L. B. and Hen, O. and Piasetzky, E. and others},
  collaboration = {CLAS and $e4\nu$ Collaborations},
  journal       = {Nature},
  volume        = {599},
  number        = {7886},
  pages         = {565--570},
  year          = {2021},
  month         = {Nov},
  publisher     = {Springer Science and Business Media LLC},
  doi           = {10.1038/s41586-021-04046-5},
  url           = {https://doi.org/10.1038/s41586-021-04046-5}
}

@article{PhysRevD.107.033007,
  title = {Introducing a novel event generator for electron-nucleus and neutrino-nucleus scattering},
  author = {Isaacson, Joshua and Jay, William I. and Lovato, Alessandro and Machado, Pedro A. N. and Rocco, Noemi},
  journal = {Phys. Rev. D},
  volume = {107},
  issue = {3},
  pages = {033007},
  numpages = {21},
  year = {2023},
  month = {Feb},
  publisher = {American Physical Society},
  doi = {10.1103/PhysRevD.107.033007},
  url = {https://link.aps.org/doi/10.1103/PhysRevD.107.033007}
}

@article{Golan:2012wx,
    author = "Golan, Tomasz and Juszczak, Cezary and Sobczyk, Jan T.",
    title = "{Final State Interactions Effects in Neutrino-Nucleus Interactions}",
    eprint = "1202.4197",
    archivePrefix = "arXiv",
    primaryClass = "nucl-th",
    reportNumber = "FERMILAB-PUB-12-280-E",
    doi = "10.1103/PhysRevC.86.015505",
    journal = "Phys. Rev. C",
    volume = "86",
    pages = "015505",
    year = "2012"
}

@article{Juszczak:2005wk,
    author = "Juszczak, C. and Nowak, J. A. and Sobczyk, J. T.",
    title = "{Spectrum of recoil nucleons in quasi-elastic neutrino nucleus interactions}",
    doi = "10.1140/epjc/s2004-02086-9",
    journal = "Eur. Phys. J. C",
    volume = "39",
    pages = "195--200",
    year = "2005"
}

@article{Prog.Part.Nucl.Phys.100.1,
title = {{NuSTEC} White Paper: Status and challenges of neutrino--nucleus scattering},
journal = {Prog. Part. Nucl. Phys.},
volume = {100},
pages = {1--68},
year = {2018},
issn = {0146-6410},
doi = {10.1016/j.ppnp.2018.01.006},
url = {https://www.sciencedirect.com/science/article/pii/S0146641018300061},
author = {L. Alvarez-Ruso and M. Sajjad Athar and M.B. Barbaro and others}
}

@article{Prog.Theor.Exp.Phys.2015.053C02,
    author = {Abe, K. and Aihara, H. and Andreopoulos, C. and others},
    collaboration = {Hyper-Kamiokande Proto-Collaboration},
    title = {Physics potential of a long-baseline neutrino oscillation experiment using a {J-PARC} neutrino beam and {Hyper-Kamiokande}},
    journal = {Prog. Theor. Exp. Phys.},
    volume = {2015},
    number = {5},
    pages = {053C02},
    year = {2015},
    month = {05},
    issn = {2050-3911},
    doi = {10.1093/ptep/ptv061},
    url = {https://doi.org/10.1093/ptep/ptv061},
    eprint = {1502.05199},
}

@article{JINST.15.T08008,
doi = {10.1088/1748-0221/15/08/T08008},
url = {https://doi.org/10.1088/1748-0221/15/08/T08008},
year = {2020},
month = {aug},
volume = {15},
number = {08},
pages = {T08008},
author = {Abi, B. and Acciarri, R. and Acero, M.A. and others},
collaboration = {DUNE Collaboration},
title = {Deep Underground Neutrino Experiment ({DUNE}), Far Detector Technical Design Report, Volume {I}: Introduction to {DUNE}},
journal = {JINST},
}

@misc{arXiv.2002.03005,
      title={Deep Underground Neutrino Experiment ({DUNE}), Far Detector Technical Design Report, Volume {II}: {DUNE} Physics},
      author={B. Abi and R. Acciarri and Mario A. Acero and others},
      collaboration={DUNE Collaboration},
      year={2020},
      eprint={2002.03005},
      archivePrefix={arXiv},
      primaryClass={hep-ex},
      url={https://arxiv.org/abs/2002.03005},
}

@article{Nucl.Instrum.Meth.A.614.87,
title = {The GENIE neutrino Monte Carlo generator},
journal = {Nuclear Instruments and Methods in Physics Research Section A: Accelerators, Spectrometers, Detectors and Associated Equipment},
volume = {614},
number = {1},
pages = {87-104},
year = {2010},
issn = {0168-9002},
doi = {https://doi.org/10.1016/j.nima.2009.12.009},
url = {https://www.sciencedirect.com/science/article/pii/S0168900209023043},
author = {C. Andreopoulos and A. Bell and D. Bhattacharya and others},
}

@article{Hayato,
	author = {Hayato, Yoshinari and Pickering, Luke},
	date = {2021/12/01},
	doi = {10.1140/epjs/s11734-021-00287-7},
	id = {Hayato2021},
	isbn = {1951-6401},
	journal = {The European Physical Journal Special Topics},
	number = {24},
	pages = {4469--4481},
	title = {The NEUT neutrino interaction simulation program library},
	url = {https://doi.org/10.1140/epjs/s11734-021-00287-7},
	volume = {230},
	year = {2021}}

@article{Phys.Rept.512.1,
title = {Transport-theoretical description of nuclear reactions},
journal = {Physics Reports},
volume = {512},
number = {1},
pages = {1-124},
year = {2012},
note = {Transport-theoretical Description of Nuclear Reactions},
issn = {0370-1573},
doi = {https://doi.org/10.1016/j.physrep.2011.12.001},
url = {https://www.sciencedirect.com/science/article/pii/S0370157311003619},
author = {O. Buss and T. Gaitanos and K. Gallmeister and H. {van Hees} and M. Kaskulov and O. Lalakulich and A.B. Larionov and T. Leitner and J. Weil and U. Mosel},
}

@article{PhysRevD.111.036032,
  title = {New multinucleon knockout model in the {NuWro} Monte Carlo generator},
  author = {Prasad, Hemant and Sobczyk, Jan T. and Ankowski, Artur M. and Bonilla, J. Luis and Banerjee, Rwik Dharmapal and Graczyk, Krzysztof M. and Kowal, Beata E.},
  journal = {Phys. Rev. D},
  volume = {111},
  issue = {3},
  pages = {036032},
  numpages = {17},
  year = {2025},
  month = {Feb},
  publisher = {American Physical Society},
  doi = {10.1103/PhysRevD.111.036032},
  url = {https://link.aps.org/doi/10.1103/PhysRevD.111.036032}
}

@article{PhysRevC.45.791,
  title = {Nuclear transparency to intermediate-energy nucleons from (e,e'p) reactions},
  author = {Pandharipande, V. R. and Pieper, Steven C.},
  journal = {Phys. Rev. C},
  volume = {45},
  issue = {2},
  pages = {791--798},
  numpages = {0},
  year = {1992},
  month = {Feb},
  publisher = {American Physical Society},
  doi = {10.1103/PhysRevC.45.791},
  url = {https://link.aps.org/doi/10.1103/PhysRevC.45.791}
}

@article{Rev.Mod.Phys.80.189,
  title = {Inclusive quasielastic electron-nucleus scattering},
  author = {Benhar, Omar and Day, Donal and Sick, Ingo},
  journal = {Rev. Mod. Phys.},
  volume = {80},
  issue = {1},
  pages = {189},
  year = {2008},
  publisher = {American Physical Society},
  doi = {10.1103/RevModPhys.80.189},
  url = {https://link.aps.org/doi/10.1103/RevModPhys.80.189}
}

@article{Phys.Rev.D.109.033008,
  title = {Lepton-induced reactions on nuclei in a wide kinematical regime},
  author = {Mosel, U. and Gallmeister, K.},
  journal = {Phys. Rev. D},
  volume = {109},
  issue = {3},
  pages = {033008},
  year = {2024},
  publisher = {American Physical Society},
  doi = {10.1103/PhysRevD.109.033008},
  url = {https://link.aps.org/doi/10.1103/PhysRevD.109.033008}
}

@article{Phys.Rev.Lett.116.192501,
  title = {Unified Description of Electron-Nucleus Scattering within the Spectral Function Formalism},
  author = {Rocco, Noemi and Lovato, Alessandro and Benhar, Omar},
  journal = {Phys. Rev. Lett.},
  volume = {116},
  issue = {19},
  pages = {192501},
  year = {2016},
  publisher = {American Physical Society},
  doi = {10.1103/PhysRevLett.116.192501},
  url = {https://link.aps.org/doi/10.1103/PhysRevLett.116.192501}
}

@article{Sealock:1989nx,
    author = "Sealock, R. M. and others",
    title = "{Electroexcitation of the $\Delta(1232)$ in Nuclei}",
    doi = "10.1103/PhysRevLett.62.1350",
    journal = "Phys. Rev. Lett.",
    volume = "62",
    pages = "1350--1353",
    year = "1989"
}

@ARTICLE{zhuang2020comprehensivesurveytransferlearning,
  author={Zhuang, Fuzhen and Qi, Zhiyuan and Duan, Keyu and Xi, Dongbo and Zhu, Yongchun and Zhu, Hengshu and Xiong, Hui and He, Qing},
  journal={Proceedings of the IEEE}, 
  title={A Comprehensive Survey on Transfer Learning}, 
  year={2021},
  volume={109},
  number={1},
  pages={43-76},
  doi={10.1109/JPROC.2020.3004555}}

@article{J.Big.Data.3.9,
  title     = {A survey of transfer learning},
  author    = {Weiss, Karl and Khoshgoftaar, Taghi M. and Wang, DingDing},
  journal   = {J. Big Data},
  volume    = {3},
  number    = {1},
  pages     = {9},
  year      = {2016},
  month     = {May},
  publisher = {Springer Science and Business Media LLC},
  issn      = {2196-1115},
  doi       = {10.1186/s40537-016-0043-6},
  url       = {https://doi.org/10.1186/s40537-016-0043-6}
}

@article{Martini:2009uj,
  author        = {Martini, M. and Ericson, M. and Chanfray, G. and Marteau, J.},
  title         = {{A Unified approach for nucleon knock-out, coherent and incoherent pion production in neutrino interactions with nuclei}},
  eprint        = {0910.2622},
  archiveprefix = {arXiv},
  primaryclass  = {nucl-th},
  reportnumber  = {LYCEN-2009-10, CERN-PH-TH-2009-192},
  doi           = {10.1103/PhysRevC.80.065501},
  journal       = {Phys. Rev. C},
  volume        = {80},
  pages         = {065501},
  year          = {2009}
}

@article{Phys.Rev.D.76.033005,
    author = "Hernandez, E. and Nieves, J. and Valverde, M.",
    title = "{Weak Pion Production off the Nucleon}",
    eprint = "hep-ph/0701149",
    archivePrefix = "arXiv",
    doi = "10.1103/PhysRevD.76.033005",
    journal = "Phys. Rev. D",
    volume = "76",
    pages = "033005",
    year = "2007"
}

@article{Phys.Rev.D.90.093001,
    author = "Graczyk, Krzysztof M. and {\.Z}muda, Jakub and Sobczyk, Jan T.",
    title = "{Electroweak form factors of the {\ensuremath{\Delta}}(1232) resonance}",
    eprint = "1407.5445",
    archivePrefix = "arXiv",
    primaryClass = "hep-ph",
    doi = "10.1103/PhysRevD.90.093001",
    journal = "Phys. Rev. D",
    volume = "90",
    number = "9",
    pages = "093001",
    year = "2014"
}

@article{Phys.Rev.D.91.033005,
    author = "Ankowski, Artur M. and Benhar, Omar and Sakuda, Makoto",
    title = "{Improving the accuracy of neutrino energy reconstruction in charged-current quasielastic scattering off nuclear targets}",
    eprint = "1404.5687",
    archivePrefix = "arXiv",
    primaryClass = "nucl-th",
    doi = "10.1103/PhysRevD.91.033005",
    journal = "Phys. Rev. D",
    volume = "91",
    number = "3",
    pages = "033005",
    year = "2015"
}

@article{Phys.Rev.C.110.054612,
    author = "Ankowski, Artur M. and Benhar, Omar and Sakuda, Makoto",
    title = "{Determination of the proton spectral function of $^{12}$C from $(e,e'p)$ data}",
    eprint = "2407.18226",
    archivePrefix = "arXiv",
    primaryClass = "nucl-th",
    doi = "10.1103/PhysRevC.110.054612",
    journal = "Phys. Rev. C",
    volume = "110",
    number = "5",
    pages = "054612",
    year = "2024"
}

@article{Phys.Rept.3.261,
    author = "Llewellyn Smith, C. H.",
    title = "{Neutrino Reactions at Accelerator Energies}",
    reportNumber = "SLAC-PUB-0958",
    doi = "10.1016/0370-1573(72)90010-5",
    journal = "Phys. Rept.",
    volume = "3",
    pages = "261--379",
    year = "1972"
}

@article{Nucl.Phys.B.Proc.Suppl.159.127,
    author = "Bradford, R. and Bodek, A. and Budd, Howard Scott and Arrington, J.",
    title = "{A New parameterization of the nucleon elastic form-factors}",
    eprint = "hep-ex/0602017",
    archivePrefix = "arXiv",
    doi = "10.1016/j.nuclphysbps.2006.08.028",
    journal = "Nucl. Phys. B Proc. Suppl.",
    volume = "159",
    pages = "127--132",
    year = "2006"
}

@article{Phys.Rev.C.53.2248,
    author = "Arrington, J. and others",
    title = "{Inclusive electron scattering from nuclei at $x \approx 1$}",
    eprint = "nucl-ex/9504003",
    archivePrefix = "arXiv",
    doi = "10.1103/PhysRevC.53.2248",
    journal = "Phys. Rev. C",
    volume = "53",
    pages = "2248--2251",
    year = "1996"
}

@article{Phys.Rev.Lett.82.2056,
    author = "Arrington, J. and others",
    title = "{Inclusive electron-nucleus scattering at large momentum transfer}",
    eprint = "nucl-ex/9811008",
    archivePrefix = "arXiv",
    doi = "10.1103/PhysRevLett.82.2056",
    journal = "Phys. Rev. Lett.",
    volume = "82",
    pages = "2056--2059",
    year = "1999"
}

@techreport{Bagdasaryan:1988hp,
    author = "Bagdasaryan, D. S. and Boyadzhian, M. K. and Kazarian, G. B. and Kechian, K. P. A. and Markarian, E. R. and Mkrtchian, G. G. and Petrosian, O. P. and Troshenkova, I. A. and Tatevosian, V. O.",
    title = "{Measurement of the spectra of $(e,e')$ scattering on $^{9}$Be and $^{12}$C nuclei in the inelastic region at $Q^2 \lesssim 0.4$~GeV$^2$/c$^2$}",
    institution = "Yerevan Physics Institute",
    number = "YERPHI-1077-40-88",
    note = "unpublished",
    year = "1988"
}

@article{Phys.Rev.Lett.61.400,
    author = "Baran, D. T. and others",
    title = "{$\Delta$ Electroproduction and Inelastic Charge Scattering From Carbon and Iron}",
    doi = "10.1103/PhysRevLett.61.400",
    journal = "Phys. Rev. Lett.",
    volume = "61",
    pages = "400--403",
    year = "1988"
}

@article{Nucl.Phys.A.402.515,
    author = "Barreau, P. and others",
    title = "{Deep Inelastic electron Scattering from Carbon}",
    doi = "10.1016/0375-9474(83)90217-8",
    journal = "Nucl. Phys. A",
    volume = "402",
    pages = "515--540",
    year = "1983"
}

@article{Phys.Rev.C.98.014617,
    author = "Dai, H. and others",
    collaboration = "Jefferson Lab Hall A",
    title = "{First Measurement of the Ti$(e,e^\prime){\rm X}$ Cross Section at Jefferson Lab}",
    eprint = "1803.01910",
    archivePrefix = "arXiv",
    primaryClass = "nucl-ex",
    doi = "10.1103/PhysRevC.98.014617",
    journal = "Phys. Rev. C",
    volume = "98",
    number = "1",
    pages = "014617",
    year = "2018"
}

@article{Phys.Rev.C.48.1849,
    author = "Day, D. B. and others",
    title = "{Inclusive electron nucleus scattering at high momentum transfer}",
    doi = "10.1103/PhysRevC.48.1849",
    journal = "Phys. Rev. C",
    volume = "48",
    pages = "1849--1863",
    year = "1993"
}

@article{Phys.Rev.Lett.105.212502,
    author = "Fomin, N. and others",
    title = "{Scaling of the $F_2$ structure function in nuclei and quark distributions at $x>1$}",
    eprint = "1008.2713",
    archivePrefix = "arXiv",
    primaryClass = "nucl-ex",
    doi = "10.1103/PhysRevLett.105.212502",
    journal = "Phys. Rev. Lett.",
    volume = "105",
    pages = "212502",
    year = "2010"
}

@article{Phys.Rev.C.35.1063,
    author = "O'Connell, J. S. and others",
    title = "{Electromagnetic Excitation of the $\Delta$ Resonance in Nuclei}",
    doi = "10.1103/PhysRevC.35.1063",
    journal = "Phys. Rev. C",
    volume = "35",
    pages = "1063",
    year = "1987"
}

@article{Phys.Rev.C.9.2230,
    author = "Whitney, R. R. and Sick, I. and Ficenec, J. R. and Kephart, R. D. and Trower, W. P.",
    title = "{Quasielastic electron Scattering}",
    doi = "10.1103/PhysRevC.9.2230",
    journal = "Phys. Rev. C",
    volume = "9",
    pages = "2230",
    year = "1974"
}

@article{Few.Body.Syst.65.78,
    author = "Mihovilovi{\v{c}}, M. and others",
    title = "{Measurement of the $^{12}$C$(e,e')$ Cross Sections at $Q^2=0.8\,$GeV$^2/c^2$}",
    eprint = "2406.16059",
    archivePrefix = "arXiv",
    primaryClass = "nucl-ex",
    doi = "10.1007/s00601-024-01944-y",
    journal = "Few Body Syst.",
    volume = "65",
    number = "3",
    pages = "78",
    year = "2024"
}

@article{Phys.Rev.D.49.4348,
    author = "Gomez, J. and others",
    title = "{Measurement of the A-dependence of deep inelastic electron scattering}",
    doi = "10.1103/PhysRevD.49.4348",
    journal = "Phys. Rev. D",
    volume = "49",
    pages = "4348--4372",
    year = "1994"
}

@article{Phys.Rev.C.107.065501,
  author        = {Al Hammal, O. and Martini, M. and Frontera-Pons, J. and Nguyen, T. H. and P{\'e}rez-Ramos, R.},
  title         = {{Neural network predictions of inclusive electron-nucleus cross sections}},
  eprint        = {2305.08217},
  archiveprefix = {arXiv},
  primaryclass  = {nucl-th},
  doi           = {10.1103/PhysRevC.107.065501},
  journal       = {Phys. Rev. C},
  volume        = {107},
  number        = {6},
  pages         = {065501},
  year          = {2023}
}

@article{PhysRevD.103.113003,
  title         = {Inclusive electron scattering and the genie neutrino event generator},
  author        = {Papadopoulou, A. and Ashkenazi, A. and Gardiner, S. and Betancourt, M. and Dytman, S. and Weinstein, L. B. and Piasetzky, E. and Hauenstein, F. and Khachatryan, M. and Dolan, S. and Megias, G. D. and Hen, O.},
  collaboration = {$e4\ensuremath{\nu}$ Collaboration},
  journal       = {Phys. Rev. D},
  volume        = {103},
  issue         = {11},
  pages         = {113003},
  numpages      = {13},
  year          = {2021},
  month         = {Jun},
  publisher     = {American Physical Society},
  doi           = {10.1103/PhysRevD.103.113003},
  url           = {https://link.aps.org/doi/10.1103/PhysRevD.103.113003}
}

@article{wreo32248,
  author   = {McElwee, Jordan},
  title    = {Tuning neutrino interaction models to inclusive electron scattering data ({PhD} Thesis)},
  journal  = {University of Sheffield},
  year     = {2022},
  month    = {September},
  url      = {https://etheses.whiterose.ac.uk/id/eprint/32248/}
}

@article{PhysRevD.111.033006,
  title     = {Implementation and investigation of electron-nucleus scattering in the neut neutrino event generator},
  author    = {Abe, Seisho},
  journal   = {Phys. Rev. D},
  volume    = {111},
  issue     = {3},
  pages     = {033006},
  numpages  = {17},
  year      = {2025},
  month     = {Feb},
  publisher = {American Physical Society},
  doi       = {10.1103/PhysRevD.111.033006},
  url       = {https://link.aps.org/doi/10.1103/PhysRevD.111.033006}
}

@article{Eur.Phys.J.C.79.293,
  author        = {Bodek, Arie and Cai, Tejin},
  title         = {{Removal Energies and Final State Interaction in Lepton Nucleus Scattering}},
  eprint        = {1801.07975},
  archiveprefix = {arXiv},
  primaryclass  = {nucl-th},
  doi           = {10.1140/epjc/s10052-019-6750-3},
  journal       = {Eur. Phys. J. C},
  volume        = {79},
  number        = {4},
  pages         = {293},
  year          = {2019}
}

@article{Phys.Rev.C.100.045503,
  author        = {Rocco, Noemi and Nakamura, Satoshi X. and Lee, T. S. H. and Lovato, Alessandro},
  title         = {{Electroweak Pion-Production on Nuclei within the Extended Factorization Scheme}},
  eprint        = {1907.01093},
  archiveprefix = {arXiv},
  primaryclass  = {nucl-th},
  reportnumber  = {FERMILAB-PUB-19-305-T},
  doi           = {10.1103/PhysRevC.100.045503},
  journal       = {Phys. Rev. C},
  volume        = {100},
  number        = {4},
  pages         = {045503},
  year          = {2019}
}

@article{AIP.Conf.Proc.1680.020013,
    author = "Zmuda, Jakub and Graczyk, Krzysztof",
    title = "{Electroweak single pion production and form factors of the $\Delta$(1232) resonance}",
    eprint = "1501.03086",
    archivePrefix = "arXiv",
    primaryClass = "hep-ph",
    doi = "10.1063/1.4931872",
    journal = "AIP Conf. Proc.",
    volume = "1680",
    number = "1",
    pages = "020013",
    year = "2015"
}

@article{Phys.Rev.D.71.074003,
    author = "Lalakulich, Olga and Paschos, Emmanuel A.",
    title = "{Resonance production by neutrinos. I. J = 3/2 resonances}",
    eprint = "hep-ph/0501109",
    archivePrefix = "arXiv",
    doi = "10.1103/PhysRevD.71.074003",
    journal = "Phys. Rev. D",
    volume = "71",
    pages = "074003",
    year = "2005"
}

@article{bslk-gxr7,
  title = {Machine learning neutrino-nucleus cross sections},
  author = {Hackett, Daniel C. and Isaacson, Joshua and Li, Shirley Weishi and Tame-Narvaez, Karla and Wagman, Michael L.},
  journal = {Phys. Rev. D},
  volume = {112},
  issue = {11},
  pages = {113005},
  numpages = {24},
  year = {2025},
  month = {Dec},
  publisher = {American Physical Society},
  doi = {10.1103/bslk-gxr7},
  url = {https://link.aps.org/doi/10.1103/bslk-gxr7}
}

\end{document}